\documentstyle[aps,epsfig,preprint]{revtex}

\newcommand{\Ree}{\mbox{$R_{e}$}}
\newcommand{\Rnn}{\mbox{$R_{n}$}}
\newcommand{\Rgg}{\mbox{$R_{g}$}}
\newcommand{\Resq}{\mbox{$\langle R_{e}^2 \rangle $}}
\newcommand{\Rnsq}{\mbox{$\langle R_{n}^2 \rangle $}}
\newcommand{\Rgsq}{\mbox{$\langle R_{g}^2 \rangle $}}
\newcommand{\Area}{\mbox{$\langle |a| \rangle $}}

\newcommand{\xikuhn}{\mbox{$k$}}
\newcommand{\gkuhn}{\mbox{$g_k$}}
\newcommand{\df}{\mbox{$d_f$}}
\newcommand{\dtop}{\mbox{$d_{t}$}}
\newcommand{\gtop}{\mbox{$g_{t}$}}

\newcommand{\dr}{\mbox{$\delta r$}}
\newcommand{\drstar}{\mbox{$\delta r^*$}}
\newcommand{\nrover}{\mbox{$p$}}
\newcommand{\phistar}{\mbox{$\phi^*$}}
\newcommand{\costh}{\mbox{$\langle\cos(\Theta)\rangle$}}

\newcommand{\beR}{\mbox{$b_e^R$}}
\newcommand{\beL}{\mbox{$b_e^L$}}
\newcommand{\bgR}{\mbox{$b_g^R$}}
\newcommand{\bgL}{\mbox{$b_g^L$}}

\newcommand{\DN}{\mbox{$D_{N}$}}

\input epsf
\begin{document}

\title{Topological effects in ring polymers (II):\\
Influence of persistence length}

\author{M. M\"{u}ller$^{1,2,*}$, J.P. Wittmer$^{2,3}$
\thanks{to whom correspondence should be addressed}
and M.E.~Cates$^2$}
\address{
$^1$ Institut f\"ur Physik, Johannes-Gutenberg Universit\"at,
55099 Mainz, Germany\\
$^2$  Dept. of Physics \& Astronomy, University of Edinburgh, \\
      King's Building, Mayfield Road, Edinburgh EH9 3JZ, UK\\
$^3$  D\'ept. de Physique des Mat\'eriaux, Universit\'e Claude Bernard
\& CNRS, 69622 Villeurbanne, France }
\date{\today}
\setcounter{page}{1}
\maketitle

\begin{abstract}
The interplay of topological constraints and persistence length of ring 
polymers in their own melt is investigated by means of dynamical Monte Carlo 
simulations of a three dimensional lattice model.
We ask if the results are consistent with an asymptotically regime
where the rings behave like (compact) {\em lattice animals} in a self-consistent 
network of topological constraints imposed by neighbouring rings. 
Tuning the persistence length provides an efficient route to increase the 
ring overlap required for this mean-field picture to hold:
The {\em effective} Flory exponent for the ring size decreases down to 
$\nu \stackrel{<}{\sim} 1/3$ with increasing persistence length.
Evidence is provided for the emergence of one additional characteristic 
length scale $\dtop \propto N^0$, only weakly dependent on the persistence 
length and much larger than the excluded volume screening length $\xi$. 
At distances larger than \dtop\ the conformational properties of the rings are 
governed by the topological interactions, at smaller distances rings and their 
linear chain counterparts become similar. 
(At distances smaller than $\xi$ both architectures are identical.)
However, the crossover between both limits is intricate and broad as a detailed discussion 
of the local fractal dimension (e.g., obtained from the static structure factor) reveals.
This is due to various crossover effects which we are unable to separate 
even for the largest ring size ($N=1024$) presented here. 
The increased topological interactions also influence the dynamical properties.
Mean-square displacements and their distributions depend crucially on the 
ring overlap and show evidence for the existence of additional size and time 
scales. The diffusion constant of the rings goes down from effectively
$\DN \propto N^{-1.22}$ for flexible rings with low overlap to 
$\DN \propto N^{-1.68}$ for strongly overlapping semi-flexible rings.
\end{abstract}

\vspace*{1cm}

\centerline{PACS numbers: 61.25.Hq, 61.41.+e, 83.10.Nn, 83.20.Fk}

\newpage
\section{Introduction}\label{sec:Introduction}

\begin{figure}
\centerline{
\epsfig{file=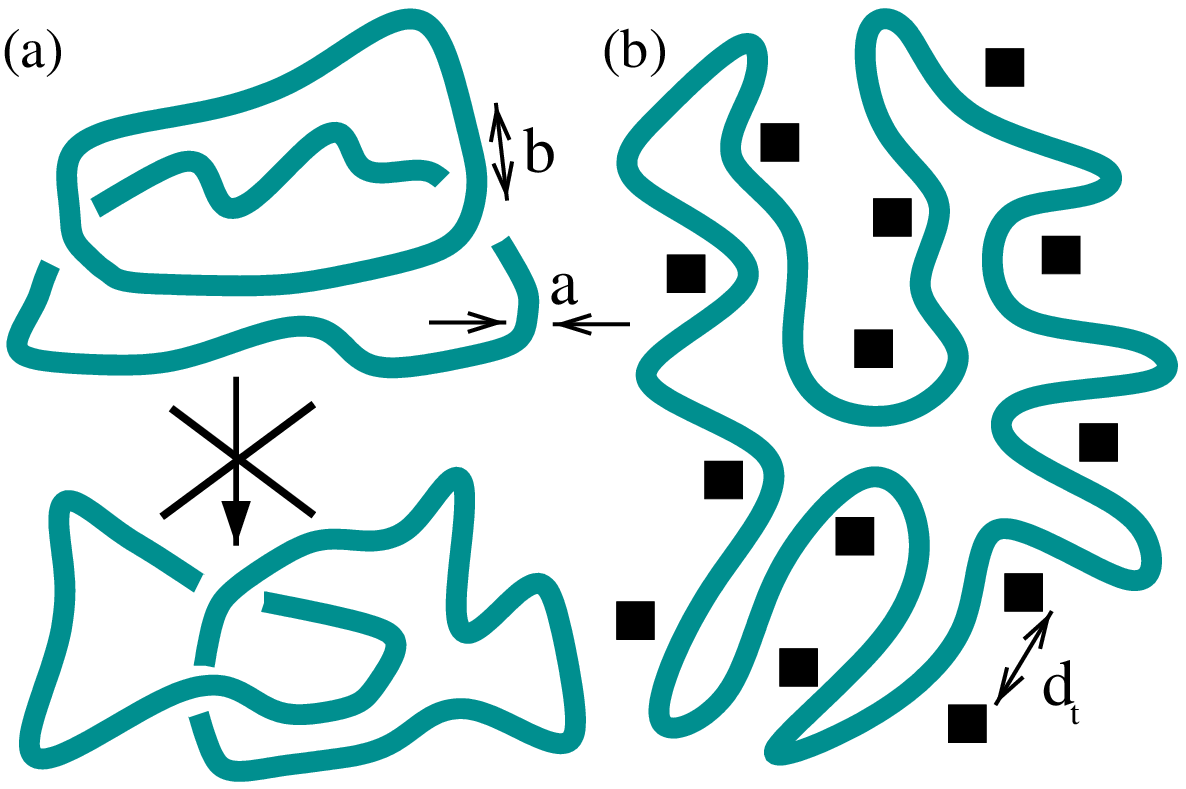,width=110mm,height=80mm}
\epsfig{file=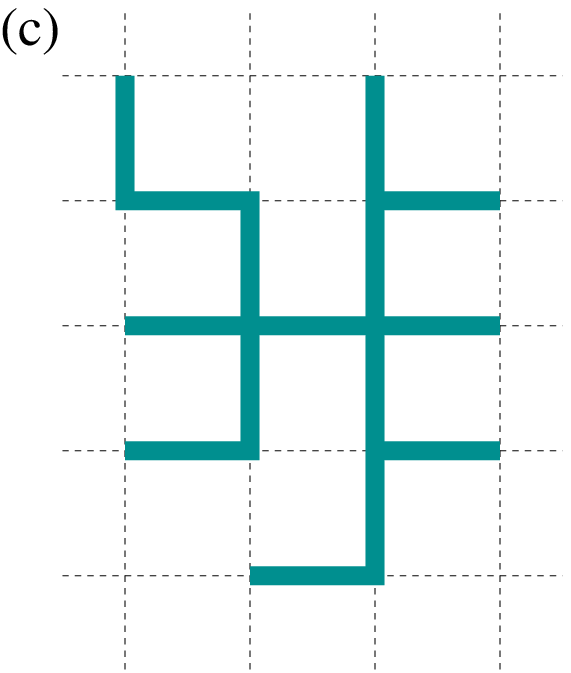,width=60mm,height=80mm}}
\vspace*{1cm}
\caption{Sketch of topological constraints and their effects.
{\bf (a)} The top chains present allowed conformations of unknotted
and (more importantly) unconcatenated rings. The rings cannot turn into
knotted configurations (not shown) or the concatenated configurations
on the bottom. It is this non-linkage constraint which tends to
{\em squeeze} a ring in semi-dilute solutions and melts.
The interplay between the topological constraint and the effective
bond length $b$ (in contrast to the monomer size $a$) is studied here.
{\bf (b)} One ring in a network of {\em fixed} topological
constraints or obstacles (squares) imposing a strongly entangled and compact 
conformation.
{\bf (c)} An equivalent Lattice Animal (LA):
In the high overlap limit rings are expected to behave effectively like LA. 
This is mean-field picture for a reference ring within a self-consistent 
topological network imposed by the surrounding rings. 
\label{fig:sketchTop}}
\end{figure}

Unconcatenated and unknotted rings in their melt are relatively 
compact\cite{com:experiments}.
This was found in recent computational studies\cite{pakula,MWC96,brown} and 
was expected on theoretical grounds\cite{CD86,ORD94,BV95,KN96}.
Qualitatively the squeezing of the rings was attributed to the topological constraints 
(see Fig.~\ref{fig:sketchTop}~(a)) \cite{com:ringclosure}. 
This is in line with much older observations showing 
that dilute rings repel each other much more strongly than their linear chain
counterparts due to the entropy loss associated with the unconcatenation 
constraint preventing the two rings to thread each other\cite{FLV75}.
While the usual excluded volume interaction is screened out at high chain overlap 
(i.e., if the ring size $R$ is larger than the size $\xi$ of the 
excluded volume blob) 
\cite{degennes,doi}, topological interactions are expected to dominate the conformational 
properties if the number of overlapping rings $\nrover \approx R^3 \phi/N$
($N$ being the chain mass, $\phi$ the monomer density) becomes high enough.

In a simple Flory-like argument Cates and Deutsch (CD)\cite{CD86} argued that 
the number of degrees of freedom lost for a typical ring due to its topological 
interactions with neighbouring rings increases like $\nrover^\alpha$ where
$\alpha$ is an unknown exponent. 
This free energy term, which favors decreasing the ring size,
has to be balanced by the entropy penalty for squashing a ring. 
To be specific, 
%the topological term was assumed to vary linearly with \nrover\ and 
this penalty was assumed to be the same as for a Gaussian chain.
Adding these contributions and minimising over $R \propto N^\nu$ yields 
the Flory exponent $\nu=(\alpha+1)/(3\alpha+2)$. 
The simplest possible estimate for $\alpha$ is to say that that roughly
one degree of freedom is lost for each of the \nrover\ neighbours which
the ring is prevented from threading, i.e. $\alpha=1$ and hence $\nu=2/5$.
This is very close to the value $\nu \approx 0.39$ 
found by us in our previous study
%using standard finite-size extrapolation techniques
\cite{MWC96}.
Note that the overlap number $\nrover \propto N^{1/5}$ increases extremely weakly
and that the only intrinsic length scale in the CD picture is the ring size $R$ itself 
(possibly renormalised in terms of excluded volume blobs of size $\xi$).
Again this was found to be qualitatively in good agreement with the simulations 
reported in our first paper~\cite{MWC96}. 
From the CD picture one expects a similar density crossover scaling for rings
as for linear chains: The ring size $R(\phi)$ reduced by the ring size $R_0$ of
the dilute reference ring should scale with $\phi/\phistar$ where $\phistar \approx N/R_0^3$
is the crossover density. Having focused in ref.~\cite{MWC96} on {\em flexible} rings at 
one fixed density we were unable to verify this implicit scaling assumption of the CD 
picture. 

A different picture comes from the extensive studies on (isolated) 
ring polymers in gels (see Fig.~\ref{fig:sketchTop}~(b)), 
often modelled by so-called {\em lattice animals} (LA), 
depicted in Fig.~\ref{fig:sketchTop}~(c)
\cite{CD86,ORD94,KN96}. 
If the ring size $R$ becomes larger than the typical distance 
\dtop\ between the fixed topological obstacles 
(represented as squares in the figure) the rings are forced to retrace 
their paths and the fractal dimension $\df=1/\nu$ becomes 
that of a strongly branched object. 
The question is now if it is possible to use this 
well-understood model {\em via} a standard mean-field argument also for strongly 
overlapping systems of rings in their own melt. 
Are the surrounding rings able to generate (in a self-consistent manner) around a 
reference ring a fixed mesh of topological obstacles? And if so, is $\dtop \propto R$
--- then LA and CD picture would be indistinguishable from the scaling point of view ---
or does it introduce an additional ring length independent scale? 
In the latter case, obviously, the above mentioned density crossover scaling would not work.
(However, one can strictly only expect this LA picture to hold in the high
entanglement limit.)
In any case, there is a catch: While increasing 
$\nrover \propto b^3 N^{3\nu-1}$, $b$ being the persistence length, 
the effective exponent $\nu(p)$ is supposed to drop down to $1/3$ 
(and even down to $\nu=1/4$ for an ideal Gaussian LA within an intermediate range of 
ring sizes $N$) \cite{ORD94}. 
Hence, the LA picture might not be self-consistent as already stressed in ref.~\cite{CD86}. 
It is crucially the prefactor $b^3$ (and not the inefficient $N^{3\nu(p)-1}$) 
which allows the simulator to control the overlap.
(This assumes that the persistence length only weakly affects the hypothetical length 
\dtop\ which has to be checked {\em a posteriori}.).

This is the route we have taken in this study to test the scaling predictions
(rather than the variation in density $\phi$ which we plan to investigate 
in a subsequent study\cite{MWC99b}). Indeed, the persistence length $b$ turns 
out to be a very efficient way to vary the overlap number \nrover\ 
(additional computational overhead and reduced diffusion constants taken into account) 
compared to the ring mass (which we have however increased from $N=512$ to $N=1024$). 
This allows us to more severely put a test on the scaling predictions of the CD scenario
than we were able in our previous study \cite{MWC96}. In contrast to that work,
the new evidence presented here shows the emergence of (at least) one further 
length scale alongside $R$ which we identify with \dtop.
Tentatively, our data are consistent with a broad crossover towards the LA picture in the 
limit of high chain overlap which is attained by increasing $b$ so that the chains become more extended
(though more compact in the scaling sense of smaller $\nu$). 
A typical semi-flexible coil at higher overlap is presented in Fig.~\ref{fig:snap}
on the right.

Of course, the CD and LA pictures are not necessarily contradictory in that 
they might provide useful heuristic descriptions in different overlap limits. 
In the computationally important regime (depending on $b$) where the rings expel 
successfully neighbouring chains and form relatively dense coils, 
the mean-field assumption of the LA approach has to break down. 
A typical chain for low overlap is shown Fig.~\ref{fig:snap} on the left.
The CD picture is {\em a priori} a good candidate to describe this regime 
where overlap \nrover\ and topological interactions are weak. 
A possible choice for the unknown CD exponent in this regime is the limiting case
$\alpha \rightarrow 0$ and, hence, $\nu \approx 1/2$, 
i.e. the topological constraints mainly contribute {\em logarithmic} 
corrections to a closed \cite{com:ringclosure} Gaussian chains of blobs. 
Evidence for this (justifying {\em a posteriori} the squeezing term used by CD)
is presented below.
 
Our paper is arranged as follows: In the next section \ref{sec:Algo} 
we give a short synopsis of the model and simulation technique used. 
To understand the special effects linked to the unconcatenation constraint we need 
reference date at the same chain and persistence length to compare with.
This is provided in section \ref{sec:diluterings} where we review
some properties of dilute semi-flexible ring polymers. 
In the subsequent section \ref{sec:densestatics} we investigate the statistics of 
ring polymers in their melt as a function of the stiffness, comparing them
with dilute rings and dense linear chains from ref.~\cite{paul,WPB92}.
(Note that in most figures we compare features discussed subsequently in 
section \ref{sec:diluterings} and \ref{sec:densestatics}.)
Possible scaling scenarios are discussed and a detailed analysis of the local 
fractal dimension is presented. 
The following section \ref{sec:dynamics} presents briefly our first results on the 
dynamics of semi-flexible ring polymers.
We conclude with a summary of our results.

\section{Algorithm and Parameters}\label{sec:Algo}

As in our previous study we investigate the properties of unknotted and 
unconcatenated rings in the framework of the bond fluctuation model 
(BFM)\cite{bfm}. Many static and dynamic properties of linear chains are 
known for this computationally efficient, coarse grained lattice model. 
A small number of chemical repeat units is mapped onto a lattice monomer 
such that the relevant characteristics of polymers -- excluded volume and 
connectivity -- are retained. Each monomer blocks a unit cell of the three 
dimensional cubic lattice from further occupancy. Adjacent monomers along 
a polymer are connected via one of 108 allowed bond vectors of lengths 
$2,\sqrt{5},\sqrt{6},3$ and $\sqrt{10}$. Here and in the following all 
spatial distances are measured in units of the lattice spacing. 
This set of bonds allows for 87 different bond angles and, hence, results 
in a good approximation to chains in continuous space. The bond vectors are chosen 
such that the local excluded volume interactions prevent the rings from 
crossing each other during their motion. This conservation of the topology 
ensures that the rings remain neither knotted with themselves nor concatenated 
with one another during the course of the simulation. We evolve the ring 
conformations {\em via} random local monomer displacements. 

As explained in the Introduction, a crucial parameter is the overlap number \nrover, 
which increases only very weakly with $N$, but much more strongly with the 
persistence length. 
In order to tune the persistence length, we impose a simple intramolecular potential 
which favours straight bond angles: $E(\Theta) = \sigma \cos(\Theta)$. 
Here $\Theta$ denotes the complementary angle between two successive bonds and
$\sigma$ the dimensionless energy scale (setting $k_BT=1$).
This potential has been used and investigated in various studies on linear polymer 
chains at filling fraction $\phi=0.5$ of occupied lattice sites \cite{WPB92,stiff}. 
This is of importance because it is reasonable to assume that on short length scales
(for long enough chains) neither ring closure nor topological constraints are pertinent.
This statement will be corroborated later in section \ref{sec:densestatics}
when we discuss the local ring structure (Fig.~\ref{fig:df}).
Note for now that other local quantities like the mean bond length $l$, \costh\
(i.e. the mean stiffness energy per monomer)
or the acceptance rate $A$ are identical for rings and their linear counterparts.
Hence, one expects to find the same local rigidity  $b(\sigma)$ for rings as for 
linear chains where $b=\beL$ is easily obtained from the mean-square end-to-end
vector \Resq\ and the known Flory exponent of linear chains.
See Tab.~\ref{tab:dilutesigma} and Tab.~\ref{tab:densesigma} for the dilute ($\phi=0$)
and the dense limit ($\phi=0.5$) respectively. 
The systems containing dilute rings and linear chains have been simulated for reasons of 
comparison. Not surprisingly the quantities featured in Tab.~\ref{tab:dilutesigma} and 
Tab.~\ref{tab:densesigma} depend somewhat on the volume fraction \cite{paul,WPB92}.
Note that at the density $\phi=0.5$ of occupied lattice sites,
many static and dynamic features of {\em molten} linear polymer materials are
reproduced by the BFM. For example, the single chain conformations obey
Gaussian statistics down to the screening length $\xi(\sigma=0) \approx 6$ of the
excluded volume interaction obtained by the static structure factor \cite{paul}.

We have restricted ourselves to values $\sigma \leq 3$ because we found
tentative evidence for the onset of a nematic order at values $\sigma \geq 4$.
Due to the lower conformational entropy rings tend to order nematically at 
lower stiffnesses than linear molecules. (For linear chains an isotropic to 
nematic transition is found close to $\sigma \approx 6$ \cite{WPB92}.)
Moreover, increasing the stiffness also decrease the number of statistically 
independent segments, and we want to keep finite $N$ effects low, 
even for small ring lengths. 

The algorithm described above has been implemented on a massively parallel CRAY T3E 
computer\cite{T3D}. Using a 2 dimensional geometric decomposition of the simulation 
grid of linear extension $L=128$, we employ 64 T3E processors. 
The simulations involve about $40\; 000$ hours of single processor CPU time. 

As shown in Tab.~\ref{tab:dense}, this allows us to investigate ring systems
with 131072 monomers and chain masses up to $N=1024$. The data for flexible rings 
have been used as starting configurations for simulations at higher persistence length.
In most cases rings diffused at least a spatial distance of 
their radius of gyration before any conformational data were collected. 
The simulation runs were extended up to five times the relaxation time
for sampling the ring conformations.
Note that we could not meet this stringent criterion for the semi-flexible ($\sigma >0$) 
systems at our largest mass $N=1024$. The resulting data appear to be time independent,
and we presume them to be equilibrated, but they have not diffused over a radius of gyration.
These three data points ($\sigma=1,2$ and $3$) have to be taken with care.

As can be deduced from Tab.~\ref{tab:dense} a relatively small increase in 
persistence length is a very effective way to increase the overlap number 
$\nrover \propto \Ree^3/N$.
To give some numbers, increasing the stiffness parameter $\sigma$ from $0$ to $3$ 
amounts in increasing \nrover\ by a factor 3 for $N=256$ (seen in Fig.~\ref{fig:snap})
and by a factor 5 for $N=512$. To achieve a similar effect by tuning the contour length
requires an increase by two, respectively three, orders of magnitude
\cite{com:estimateAlgo}.

\section{Dilute rings revisited}\label{sec:diluterings}

As a reference for the subsequent section on strongly overlapping rings
(where the unconcatenation constraint matters) we briefly revisit some properties of 
flexible and {\em semi-flexible} dilute rings and their linear counterparts. 
Basically, the remaining unknottedness constraint and 
the requirement of ring closure\cite{com:ringclosure}
do not matter: Dilute rings behave broadly like linear chains.

\begin{figure}
\centerline{\epsfig{file=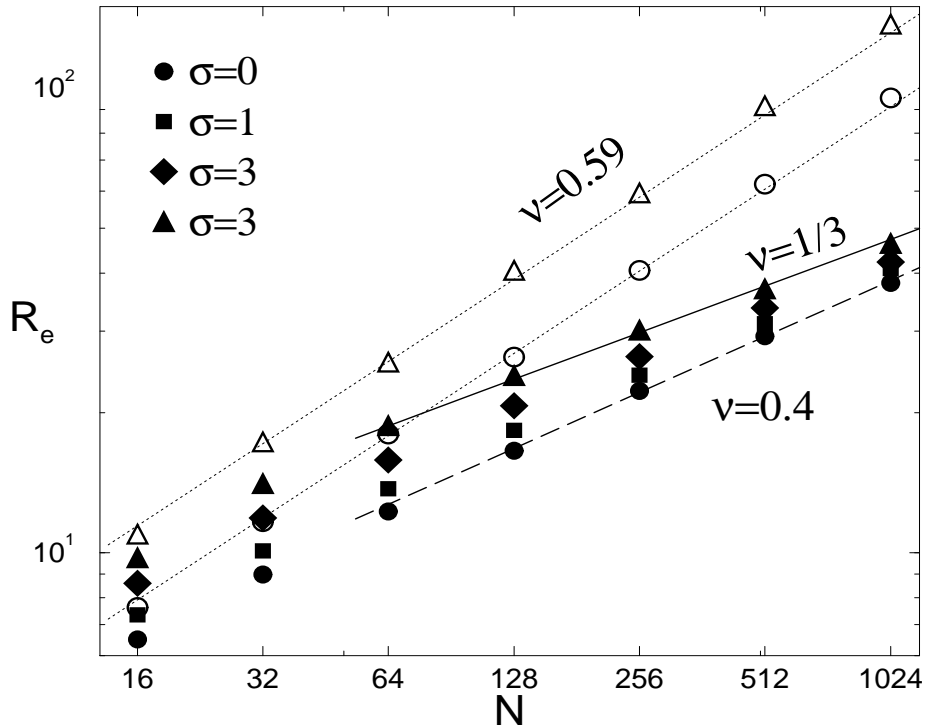,width=120mm,height=100mm}}
\vspace*{1cm}
\caption{Diameter \Ree\ versus chain length $N$ for 
different $\sigma$ for dilute (open symbols) and dense 
(full symbols) rings. 
The dilute rings are characterized by the same Flory
exponent $\nu = \nu_0 \approx 0.59$ as their linear counterparts (not shown).
For large enough chains this is independent of $\sigma$.
For the dense chains the effective fractal dimension $\df = 1/\nu$ 
becomes much larger, depending strongly on the chain rigidity:
flexible chains ($\sigma=0$) compare well with $\nu=0.4$ (dashed line),
semi-flexible chains ($\sigma=3$) with $\nu=1/3$ (solid line).
\label{fig:RsN}}
\end{figure}

\begin{figure}
\centerline{\epsfig{file=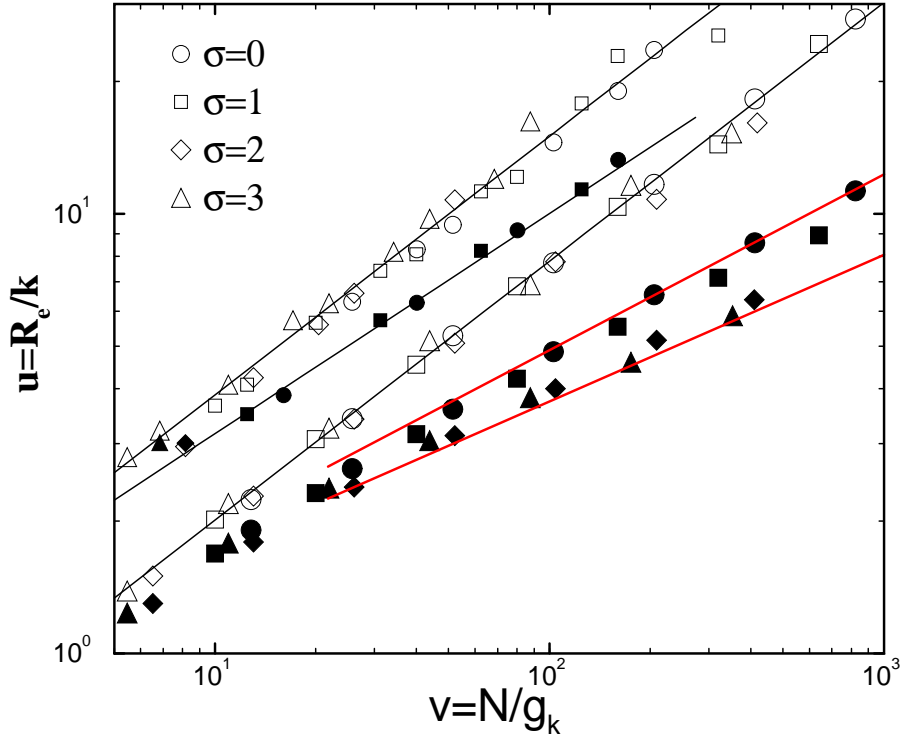,width=120mm,height=100mm}}
\vspace*{1cm}
\caption{Reduced diameter $u=\Ree/k$ versus reduced chain mass $v=N/\gkuhn$
for dilute ($\phi=0$, open symbols) and dense ($\phi=0.5$, full symbols) 
linear chains (top two lines) and rings (bottom). 
The chain flexibility $\sigma$ changes as indicated in the figure.
The lines indicate (effective) exponents (from top to bottom):
$\nu=0.59$ for dilute linear chains, 
$\nu=0.5$ for dense linear chains,
$\nu=0.59$ for dilute rings,
$\nu=0.4$ for dense, flexible ($\sigma=0$) rings and
$\nu=1/3$ for semi-flexible ($\sigma=3$) rings in the melt.
Note that the dense rings do not collapse on one master curve as their linear counterparts.
\label{fig:Rkg}}
\end{figure}

The simplest quantity to characterise the chain structure as a function of the 
stiffness parameter $\sigma$ is the overall chain size versus chain length $N$.
The size of the rings is measured first with the usual mean-square radius
of gyration \Rgsq. As a second measure, we sample the mean-square distance 
$\Rnsq=\langle (\vec{R}_i-\vec{R}_{i+n})^2 \rangle$ (averaged over all monomers $i$)
between pairs of monomers that are $n$ monomers apart along the contour.
In particular we define for rings the mean-square ring diameter 
$\Resq  = \Rnsq|_{n=N/2}$.
(For the linear chains $\Resq=\Rnsq|_{n=N}$ denotes the usual mean-square end-to-end 
distance.)
In Fig.~\ref{fig:RsN} the diameter $\Ree=\Resq^{1/2}$ for flexible 
and semi-flexible dilute rings is plotted versus $N$. 
Not included are the linear counterparts of same $N$
and $\sigma$. (But see Fig.~\ref{fig:Rkg}.) 
The data are well fitted for all the persistence lengths considered by  
$\Ree = b_e N^{\nu}$ with the classical Flory exponent 
$\nu=\nu_0 \approx 3/5$ \cite{deutsch}.
The effective bond length $b_e$ encodes the persistence length effect. 
It is tabulated for linear chains ("L") and rings ("R") in table~\ref{tab:dilutesigma}. 
Similar fits have also be performed for the radius of gyration. 
We find, independent of chain size and persistence length, the following relations:
$\beL/\beR \approx 1.89$, $\bgL/\beL \approx \sqrt{6}$ and $\beR/\bgR \approx 1.79$.
Note that in fitting we have disregarded the three smallest masses $N=16, 32$ and $64$
to minimize finite size effects. 
We have checked for (the remaining) finite size effects (not shown): 
The (negative) curvature in the data points is surprisingly weak and 
{\em similar} for linear chains and rings. 
It is slightly stronger for the radius of gyration
than for the diameter which probes larger distances. 
While we cannot rule out that the asymptotic exponents of the ring
chains are slightly different from their linear counterparts,
for all practical purposes of relevance here we may conclude that neither 
the ring-closure nor the non-selfknottedness constraint are pertinent. 

We are now in the position to rescale the effects of the persistence length for both 
linear chains and rings by estimating for the linear chain the number \gkuhn\ of correlated 
monomers along the chain and the associated Kuhn length $\xikuhn=l \gkuhn$
($l$ being the measured mean length of the BFM bond). 
Rewriting $R_e = b_e N^\nu = \xikuhn (N/\gkuhn)^\nu$ yields 
$\gkuhn=(b_e/l)^{1/(1-\nu)}$ and $\xikuhn=l (b_e/l)^{1/(1-\nu)}$.
%The corresponding values are given in tab.~\ref{tab:dilutesigma}.
Obviously this scaling-based definition of \xikuhn\ and \gkuhn\ 
is arbitrary to within prefactors of order one. 
The proposed rescaling works successfully for dilute 
linear chains and rings as shown in Fig.~\ref{fig:Rkg}.
Here we have plotted the reduced diameters $u=\Ree/\xikuhn$ 
versus the number of statistical segments $v=N/\gkuhn$.
A similar figure was obtained for the radius of gyration (not shown).
Note that only a part of the linear chain configurations used to characterise the stiffness
effects for dilute systems are included in the figure.

\begin{figure}
\centerline{\epsfig{file=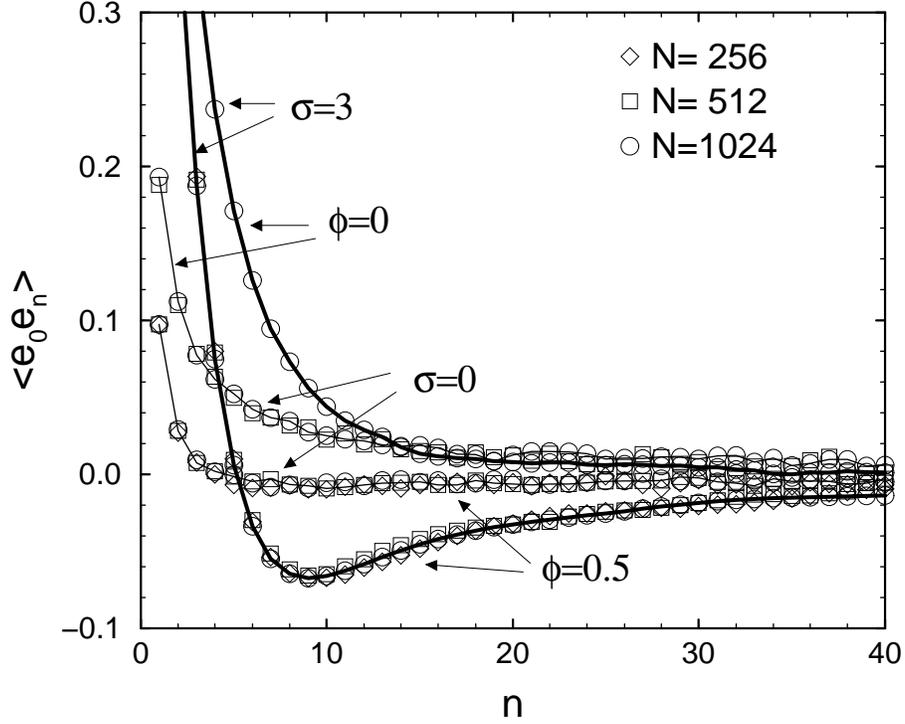,width=120mm,height=100mm}}
\vspace*{1cm}
\caption{Bond-Bond correlation function $<e_0 e_n>$ for 
dilute ($\phi=0$) and molten ($\phi=0.5$) rings for $\sigma=0$ 
(connected by thin lines) and $\sigma=3$ (connected by fat lines)
for various chain masses as indicated in the figure.
At high chain overlap (high density and rigidity) we find a pronounced negative
dip in the correlation function.
\label{fig:eiej}}
\end{figure}

In Fig.~\ref{fig:eiej} we show the bond-bond correlation function 
$<\vec{e}_i \vec{e}_{i+n}>$, where $\vec{e}_i$ denotes one of the 
bond vectors of monomer $i$ (say the ``right" one in a given list). 
Only values for rings are included (the values for linear chains being 
identical for long enough chains). 
As it should, the bond-bond correlation functions for different persistence lengths
collapse onto a single scaling curve, when plotted versus $n/\gkuhn$ (not shown). 
Note that this scaling and the (supposed) exponential decay is sometimes preferred 
to define the number of correlated monomers \gkuhn\ and the 
persistence length \xikuhn\ rather than our definition based on the measured 
effective bond length $b$ and the well-known (and for rings confirmed) Flory exponent. 
However, the correlation function for dilute chains (both linear and rings) does 
certainly {\em not} decrease as a pure exponential (not shown). This is due to 
(non-universal) short range packing effects and (more importantly) to long range 
excluded volume correlations. 
In addition to this we prefer our method for statistical reasons.

Additional information about the local structure of the flexible and
semi-flexible rings comes from the structure factor $S(q)$
which is of direct experimental relevance. The slopes of the structure 
factor in $\log-\log$ coordinates define the differential fractal dimension 
$\df(q)= - d\log(S(q))/d\log(q)$. The general shape of $S(q)$ was discussed
in ref.~\cite{MWC96}. We present here only $\df(q)$
which contains all the information. It is equivalent in the limit of 
long chains ($q\rightarrow 0$, $q \Rgg \gg 1$) 
to the inverse of the Flory exponent $\nu$.
Fig.~\ref{fig:df}(a) shows $\df(q)$ for dilute rings (thin symbols) of mass $N=1024$
plotted versus $\dr = 2\pi/q$. For large distances the rings are well described by 
the classical Guinier expansion $S(q) \approx N \left(1-(\Rgg q)^2/3 \right)$
(not shown); hence $\df(q\Rgg\rightarrow 0) \rightarrow 0$. 
For small distances $\dr \approx l$ one finds the usual spurious Bragg-peak 
of the structure factor \cite{MWC96} 
due to local packing effects giving rise to negative \df-values. 
We note a weak $\sigma$-dependence due BFM related lattice effects:
A slightly different set of bond vectors is preferred.
This is in line with the fact that also the mean bond vector varies weakly
with persistence length as shown in Tab.~\ref{tab:dilutesigma}.
In between both limits we find a broad plateau for the flexible rings 
with $\df = 1/\nu_0 \approx 1.7$ (upper horizontal line) due to the 
excluded volume interactions.
Semi-flexible rings show, as expected from linear chains, a shoulder at about $\df =1$ 
(upper horizontal line) due to the local rigidity of the rings.
Note that the curves for linear dilute chains (not included) are very similar. 
The only qualitative difference is the ``hump" between the plateau and the Guinier regime.
It is due partially to the ring closure as one can easily see by calculating the
static structure factor for a Gaussian ring (solid line).

\section{Conformational properties of dense rings}\label{sec:densestatics}

\subsection{Compact, but strongly overlapping rings}
\label{subsec:A}

While dilute rings essentially behave like their linear counterparts, this becomes very 
different for unconcatenated rings in the melt. This is shown in the dramatic decrease 
of the chain size of rings (full symbols) in Fig.~\ref{fig:RsN} as compared
with dilute reference rings (open symbols). Certainly, linear chain sizes also 
decrease with increasing density due to the screening of the excluded volume 
correlations \cite{degennes,paul}, but to nowhere near this extent,
as Fig.~\ref{fig:Rkg} illustrates.
Moreover, the slopes of \Ree\ versus $v = N/\gkuhn$ 
decrease strongly with persistence length. 
(See Tab.~\ref{tab:densesigma}.) For flexible rings ($\sigma=0$) the data points 
(for $N=128$ to $N=1024$) are fitted by an (effective) exponent $\nu=0.4$.
For our stiffest rings we observe $\nu(\sigma=3)\approx 1/3$.
This last result is indeed consistent with the compact LA scaling behaviour
predicted for rings at high enough overlap \nrover.
 
\begin{figure}
\centerline{\epsfig{file=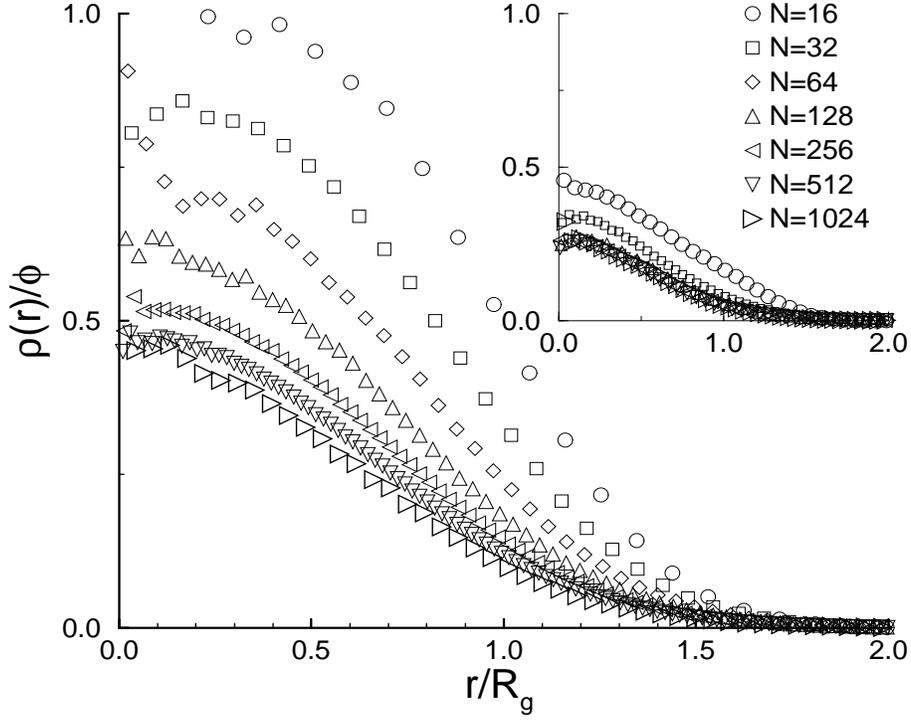,width=120mm,height=100mm}}
\vspace*{1cm}
\caption{Single chain density distribution with respect to its center of mass
for dense systems $\phi=0.5$ with flexible rings ($\sigma=0$)
versus reduced distance $r/\Rgg$.
The density at the center of the coil decreases ($\Rgg^3/N \propto N^{-0.2}$
in agreement with the effective exponent from Fig.~\ref{fig:RsN}),
but remains always larger than $0.4 \phi$.
The inset presents the same for semi-flexible rings ($\sigma=3$);
the (rescaled) distribution density becomes chain length independent
for $N \le 64$ (density at the center $\approx 0.25 \phi$).
\label{fig:gr}}
\end{figure}

These results are also in agreement with the density distribution of a ring around its 
center of mass, shown for rings in Fig.~\ref{fig:gr}. We see that for flexible rings 
(main figure) the overlap $\nrover$ must be very small; nearly all monomers from 
neighbouring rings are expelled. 
From the Flory exponent $\nu\approx 0.4$ measured above one expects the
density inside the ring to decrease as $\rho \sim N^{-0.2}$. This is confirmed.
In the inset we present the corresponding plot for semi-flexible rings ($\sigma=3$).
Since the effective Flory exponent is $\nu=1/3$ we expect the density inside 
the ring to be independent of the ring size. The ring length independence of the
density is observed for $N \stackrel{>}{\sim} 128$ for $\sigma=2$ (not shown) and 
$N \stackrel{>}{\sim} 64$ for $\sigma=3$. This supports the idea that the LA regime 
is reached earlier for stiffer rings. Though rings in this regime show compact 
scaling ($\nu\approx 1/3$) the asymptotically constant density $\rho$ is fairly small.
Thus the density of {\em other} rings in the correlation hole of a given ring is 
{\em larger} for stiff than for flexible rings (Fig.~\ref{fig:gr}).
This holds for the ring sizes studied; it does not exclude similar strong overlap behaviour for 
flexible, but much larger rings.

\subsection{First evidences for additional length scale}
\label{subsec:B}

We have to stress, however, that the above exponents are only {\em effective} exponents. 
The curves are strongly (negatively) curved as one can easily visualise by plotting 
$R/N^{1/3}$ versus $N$ (not shown). 
Alternatively, one can characterise this curvature {\em via} the differential Flory exponent 
$\nu(N)$ defined from the increase between chain length $N$ and $2N$ \cite{MWC96}. 
This reveals (not shown) that $\nu(N)$ decreases continuously from $\nu \approx 1/2$, 
i.e. nearly Gaussian behaviour, for smaller rings down to the slopes indicated 
in Fig.~\ref{fig:RsN} for the largest masses we were able to simulate. 
In our previous study \cite{MWC96} we have attributed the observed
curvature to classical finite-size effects which also appear
for linear chains. That is, we have not attributed them
to topological effects, but excluded volume interactions visible
due to the finite number of blobs. 
If one admits this as the {\em only} physical origin of the curvature 
(excluding, e.g., additional length scales) it is indeed reasonable to attempt 
to obtain the asymptotic exponent by classical finite-size extrapolation method.
In this method, the local slope $\nu(N)$ (obtained from the diameter and the radius of 
gyration) is plotted versus the reduced blob size  $\xi/R$
where $R$ is itself given in a self-consistent manner by the asymptotic
behaviour. This works generally well for linear polymers.
Proceeding along these lines we obtained an asymptotic exponent of 
$\nu \approx 0.4$ (which we now obtain directly after including $N=1024$). 
This value happened to coincide with the CD picture ($\alpha=1$ \cite{CD86})
which made us perhaps more confident than we should have been (in ref.\cite{MWC96})
of having characterised the asymptotic behaviour.

The excluded volume effects (and their screening) do certainly contribute 
to the general crossover scenario. 
This is shown below when we discuss the structure factor (Fig.~\ref{fig:df}). 
But a danger remains that we may have missed additional physics,
in the form of an additional length scale (not included in the simple CD picture),
which controls the behavior of molten rings.

\begin{figure}
\centerline{\epsfig{file=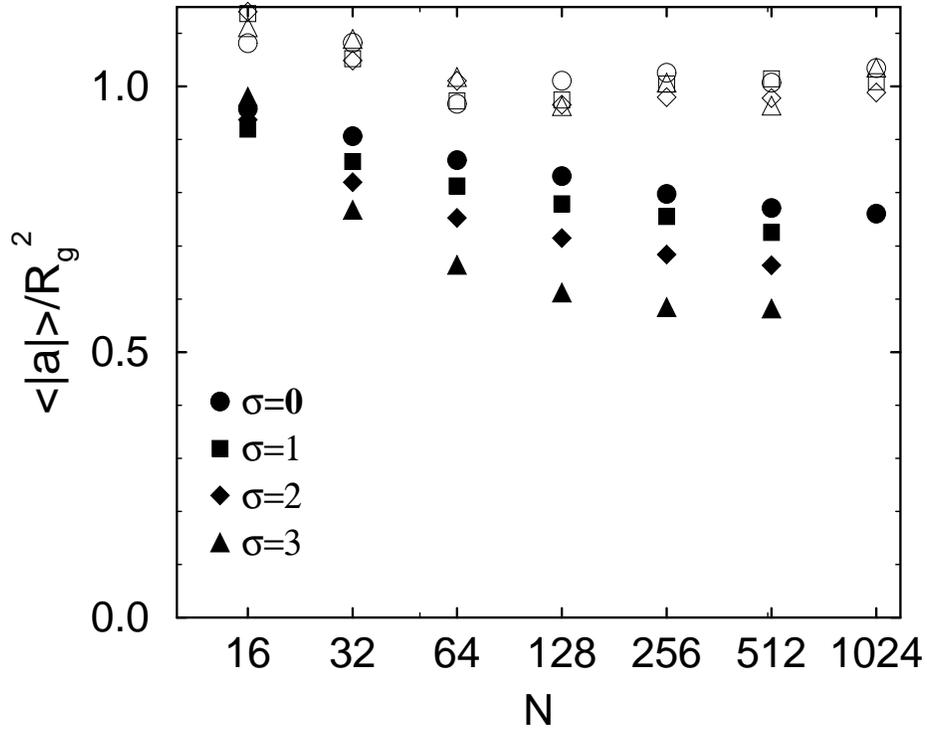,width=120mm,height=100mm}}
\vspace*{1cm}
\caption{Reduced area $\Area/\Rgsq$ versus $N$ for dilute rings (open symbols)
and molten rings (full symbols) for different $\sigma$. 
Note the logarithmic, but systematic correction to the scaling
for the latter.
\label{fig:AN}}
\end{figure}

Thus one of the key claims made in ref.~\cite{MWC96} was that the only 
relevant length scale in dense rings is the chain size itself. 
This was shown in various scaling plots, e.g.,
of the reduced mean ring ``area" $\Area/\Rgsq = const$. The area $a$ is defined as a 
signed quantity (the component in that direction of the vector area of a ring) 
that vanishes for any configuration in which the ring exactly retraces its own steps.
Qualitatively, one expects that a ring which is only very weakly threaded by the surrounding 
rings has a very small area. This should hold in particular while we approach the LA limit. 
The reduced area $\Area/\Rgsq$ has been (log-linear) plotted in Fig.~\ref{fig:AN} versus 
ring mass $N$ for dilute rings (empty symbols) and rings in the melt (full symbols) of
different flexibility. For (long enough) {\em dilute} rings the ratio is indeed chain length
(and persistence length) independent.
For the dense limit however, we now find evidence 
for (logarithmic) corrections to the expected CD scaling (plateau). 
The reduced area decreases systematically as a function of mass 
--- this is in contrast to what was deduced from a (linear-linear) plot in 
ref.~\cite{MWC96}.
Note that the ratio decreases also with respect to $\sigma$: stiffer rings are 
likely to get less threaded by other rings. This is consistent with the LA picture 
where stiffer (and hence larger) rings are forced to retrace their own steps in a network 
of fixed obstacles.

\subsection{Failure of classical scaling analysis without additional length scale}
\label{subsec:C}

\begin{figure}
\centerline{\epsfig{file=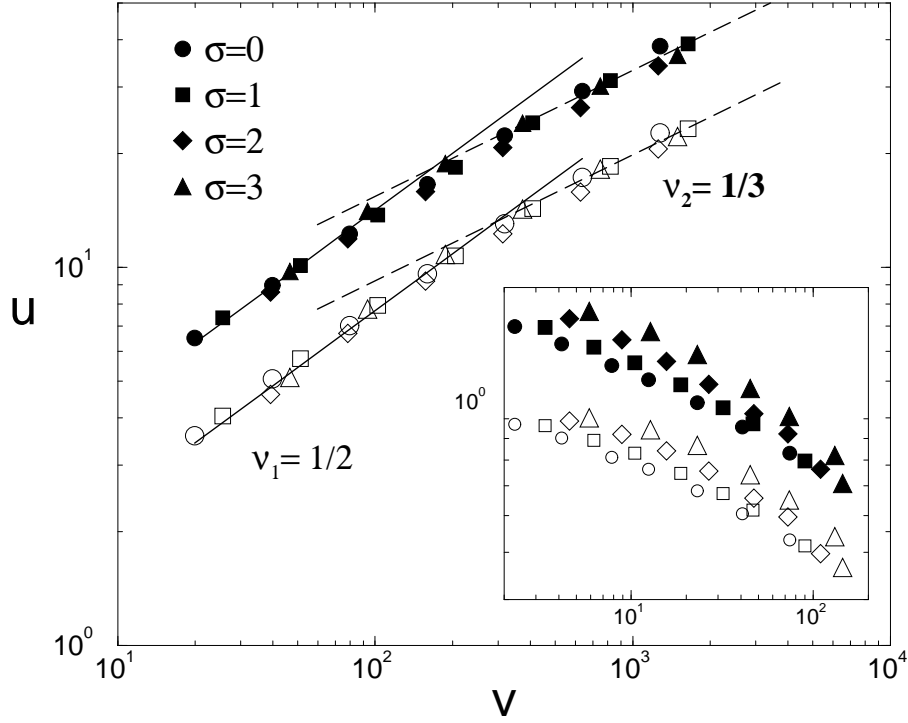,width=120mm,height=100mm}}
\vspace*{1cm}
\caption{Two scaling attempts for diameter (full symbols)
and radius of gyration (empty symbols) at different flexibilities
as indicated in the figure.
In the inset we try the traditional scaling $u=R/R_0$ versus
$v=\phi/\phistar \approx R_0^3 \phi/N$. 
The poor scaling evidences additional length scales.
In the main figure we explore the possibility of a length scale \dtop\
assumed to be independent of both mass $N$ and persistence length $\sigma$.
We plot $u=R/\dtop$ versus $v=N/\gtop$ where we set arbitrarily $\dtop=1$ 
and fix $\gtop/\gkuhn = (\dtop/\xikuhn)^2$ self-consistently with the observed effective 
Gaussian behavior at short distances. All data points collapse. 
The slopes indicate Gaussian behavior for $R < \dtop$ ($\nu_1=0.5$: solid line) 
for $R < \dtop$ and compact LA ($\nu_2=1/3$: dashed line) for larger distances.
\label{fig:uv}}
\end{figure}

More stringent tests for the CD picture are posed by the following two scaling 
analyses which try to allow for a varying persistence length under the
assumption of no new additional length scale for topological interactions. 
They are presented in Fig.~\ref{fig:Rkg} and in the inset of Fig.~\ref{fig:uv}. 

In the first we replot the reduced ring size $u=\Ree/\xikuhn$ versus the number of 
statistical units $v=N/\gkuhn$ as we did in Section~\ref{sec:diluterings} for dilute rings. 
The values of \xikuhn\ and \gkuhn\ used in the Fig.~\ref{fig:Rkg} 
for the dense linear chains and rings are the same as for the dilute systems
\cite{com:valuegauss}. 
The difference to the behaviour of linear chains is striking: The ring data for different $\sigma$-lines {\em diverges} 
while the linear chain data clearly collapses (on the expected slope $\nu=0.5$). 
This is a physically unsound result; flexibility-dependent universality classes 
for rings are difficult to accept.

The second scaling test is the ``classic" $u=R(\sigma,\phi=0.5)/R_0$
versus $v=\phi/\phistar \approx N/R_0^3$ mentioned in the Introduction \cite{degennes}.
This is shown in the inset of Fig.~\ref{fig:uv}. Both ring diameter (full symbols)
and radius of gyration (empty symbols) at $\phi=0.5$ have been included. 
As reference chain size $R_0$ we used the measured radius of gyration for a 
dilute ring of given $\sigma$. 
Again, the scaling attempt fails to account for data of more than one chain stiffness.

For the moment we may conclude that there is clear evidence for one or more
additional length scales. At this point, we still do not know how this hypothetical 
length scale \dtop\
depends on ring and persistence lengths, nor do we know what physics it represents:
however, since it arises for rings and not for linear chains, we can presume a
topological origin.

\subsection{Evidence for a chain length independent length \dtop}
\label{subsec:D}

Another striking effect which we are able to see due to the
persistence length variation is shown in Fig.~\ref{fig:eiej}
for the bond-bond correlation function of dense rings.
As before for dilute chains this correlation function becomes
chain length independent for the large masses indicated in the figure.
For flexible rings the correlation function drops down slightly below
zero within three monomers and approaches then zero from below.
When the persistence length is increased a most remarkable negative
correlation becomes visible indicating that the polymer is likely
to fold back after 10 monomers for ($\sigma=3$)\cite{com:needle}.
Needless to say that the correlation function for dilute rings
does not show anything faintly similar. 
The position and the depth of this anti-correlation dip increases with $\sigma$. 
We stress that the position and the shape of the dip are chain length independent.
If this effect has something to do with the additional length scale \dtop\ mentioned above,
as we believe, this is a piece of evidence for its chain length independence.

\subsection{A new scaling scenario}
\label{subsec:E}

\begin{figure}
\centerline{\epsfig{file=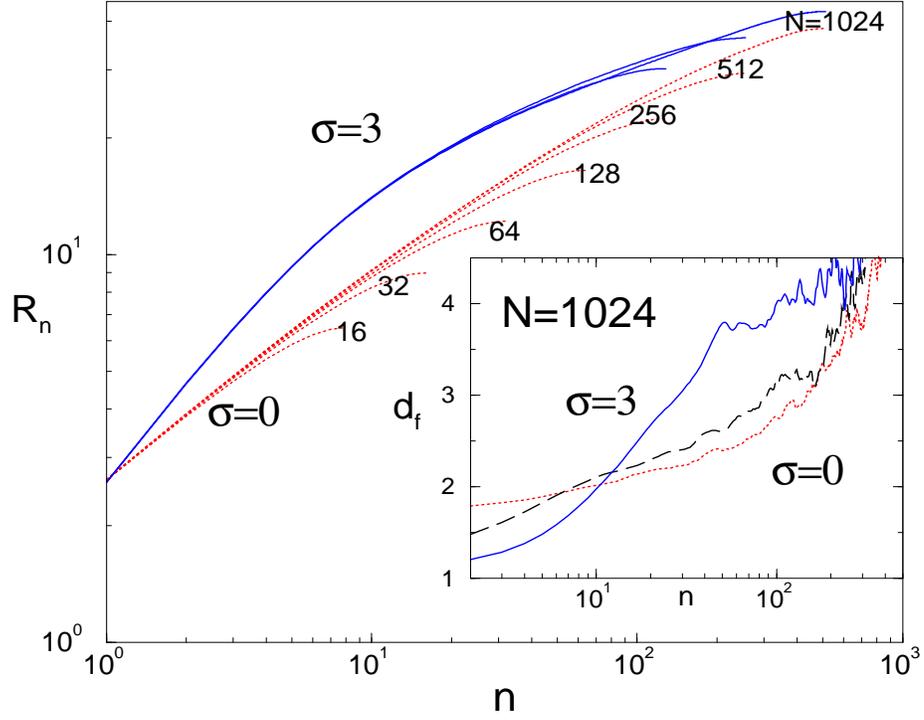,width=120mm,height=100mm}}
\vspace*{1cm}
\caption{Distance \Rnn\ between a chain contour of length $n$
for $\phi=0.5$.
Main figure: Flexible and semi-flexible ($\sigma=3$) chains
of various chain length $N$ as indicated in the figure.
Inset: Differential fractal dimension $\df(n)=1/\nu(n)$ for $N=1024$
and $\sigma=0$ (dotted line), $\sigma=1$ (dashed line)
and $\sigma=3$ (solid line).  
\label{fig:Rn}}
\end{figure}

This observation forms the basis for the next scaling proposal attempted in Fig.~\ref{fig:uv}.
We assume here that there is one additional, mass independent length scale \dtop.
Hence, the associated number \gtop\ of monomers between the topological obstacles
is also mass independent, but depends on the local conformational properties.
These are complicated as will be revealed below (see Fig.~\ref{fig:df} and Fig.~\ref{fig:Rn}). However, as we have observed above (Fig.~\ref{fig:RsN}),
short rings are reasonable approximated by Gaussian statistics and we may write 
$\gtop = \gkuhn \left( \dtop/\xikuhn \right)^{1/\nu_1}$,
where $\nu_1$ is some effective exponent close to $1/2$ characterising the (messy) 
statistics at distances $\xikuhn \ll \dr \ll \dtop$. 
We want to plot $u=R/\dtop$ versus $N/\gtop$ where $R$ denotes either the diameter 
(full symbols) or the radius of gyration (empty symbols). 
We still need to fix the persistence length dependence of \dtop.
For simplicity, we suppose that \dtop\ is only weakly affected by $\sigma$
and set arbitrarily $\dtop=1$, i.e., we neglect any stiffness dependence of \dtop\ itself.  
Hence, $\gtop \propto 1/\gkuhn$ and $v=N \gkuhn$.
This assumes that the effect of $\sigma$ is to increase $R/\dtop$
(allowing the asymptotic regime to be accessed for smaller $N$)
by increasing $R$ at (nearly) constant \dtop.
This idea is very much the opposite of what we tried in Fig.~\ref{fig:Rkg}
where we had $v=N/\gkuhn$ and stiffer chains with fewer statistical units 
were considered as being effectively ``shorter".
As shown in Fig.~\ref{fig:uv} this simple new proposal is successful although
a weak dependence of \dtop\ on $\sigma$ cannot be ruled out.
Moreover, it is self-consistent: 
The local statistics is indeed well described by the Gaussian behaviour $\nu_1 \approx 1/2$.
This might be interpreted as the $\alpha \rightarrow 0$  limit of the CD proposal.
The data collapse over two orders of magnitude in $v$ justifies {\em a posteriori} 
the neglect of a stiffness dependence of \dtop.
At large distances data points for {\em different} persistence lengths fall together on
one slope of $\nu_2=1/3$ which is consistent with the LA picture. 
Note the later crossover of the radius of gyration onto the $\nu_2$-slope.
This might be related to the fact that \Rgg\ probes smaller distances.

The above proposal may be in principle generalised to incorporate excluded volume effects 
on intermediate scales $\xikuhn \ll \dr \ll \xi$ where the statistics is governed
by the exponent $\nu_0=0.59$ for dilute rings. Note that the strong dependence of $\xi$
on the persistence length is checked by the small difference between $\nu_0$ and $\nu_1$.
(We plan to consider this problem in a subsequent study on density effects \cite{MWC99b}.) 
This offers an effective simplification for the molten chains ($\phi=0.5$) of interest in this study:
Both length scales $\xi$ and $\xikuhn$ are of same order and we are not able to separate
the effects anyway as we are going to show now. 

\subsection{Local conformational properties: Structure factor}
\label{subsec:F}

So far we have considered mainly global properties like the ring diameter \Ree\ or the
ring area. We wish now to characterise the {\em local} conformational properties
by measuring the single chain structure factor $S(q)$ 
and the mean-square length \Rnsq, 
both introduced in Sec.~\ref{sec:diluterings},
and the differential fractal dimensions $\df(q)$ associated with both quantities.
We will first consider the fractal dimension obtained from $S(q)$
as presented in Fig.~\ref{fig:df} and then compare this in Sec.~\ref{subsec:F} below) 
with \df\ obtained from \Rnsq\ and shown in Fig.~\ref{fig:Rn}.

The differential fractal dimension for our longest ($N=1024$) rings in the melt 
($\phi=0.5$) is depicted in Fig.~\ref{fig:df}(a) and compared with their dilute
counterparts. 
The Bragg peak at $\dr \approx l$ is much more pronounced than for the dilute chains
(not fully shown). We note again small, but distinct $\sigma$-dependent packing effects.
Not surprisingly, at very large distances the structure is again well 
described by 
the Guinier expansion (see section \ref{sec:diluterings}) and $\df(q)$ vanishes 
smoothly.

Obviously one wants to understand the behaviour between the (featureless and trivial) 
Guinier and the (non-universal) Bragg part. To stress this regime we have plotted in 
Fig.~\ref{fig:df}(b) the differential fractal dimension $\df(q)$ versus $\dr=2\pi/q$ 
chopping off distances $\dr < l$ and the Guinier part. 
Data are for flexible ($\sigma=0$) and semi-flexible ($\sigma=3$) rings 
and linear chains in the melt. 
For small distances where $\df < 1$ the data for all \df\ are very similar.
(Note the small $\sigma$-dependent packing effects mentioned above.)
At $\df(\dr \approx 4) = 1$ the semi-flexible systems branch off the much steeper 
line for flexible chains (top line). 
This applies also to the data for $\sigma=1$ and $\sigma=2$ (not shown). 
We stress that the difference between different $\sigma$ is gradual and 
that systems of flexible rings do not constitute a singular limit.
Because of the small blob size (i.e., the very strong interactions with the 
surrounding rings) we are unable to separate rigidity and excluded volume effects:
Semi-flexible rings do not show the ``shoulder" at 
$\df\approx 1$ we saw for dilute rings (dotted line in Fig.~\ref{fig:df}(a)).  

At these small distances, smaller than the blob size $\xi$ \cite{paul}, 
linear chains behave exactly like their ring counterparts of the same flexibility. 
This proves, as assumed above, that on small scales the $\sigma$-effects are the same 
for both architectures and justifies the use of the persistence length $b$ 
(and the associated values \gkuhn\ and \xikuhn) obtained from linear chains. 
It also shows that it was natural to attempt in ref.~\cite{MWC96}
a finite-size scheme in terms of the correlation hole, at least when one
assumes the CD picture to hold, i.e. $R$ being the only length scale.

However, the rescaling in Fig.~\ref{fig:Rkg} did {\em not} work for molten rings 
(in contrast to linear chains). This is due to effects caused by the 
ring closure and topology which intervene at larger distances $\dr \gg \xi$
(and in particularly at $\dr \gg \dtop$): 
The fractal dimension \df\ for rings continues to rise while the linear chains 
become roughly Gaussian, i.e. $\df \approx 2$ \cite{com:warninglinear}. 
The more flexible the rings, the more this increase becomes suppressed
(see the flexible chains of mass $N> 256$ in Fig.~\ref{fig:df}(b)).
This gives the more rigid systems a chance to catch up
(in $\df(q)$) the more flexible ones ---
and this after having consumed fewer monomers on short scales. 

The above discussion is only valid if the chains are actually large enough
i.e. $\dr \ll \Rgg$ or $S(q) \ll N$, 
to fall on the asymptotic ($\sigma$-dependent) curves. 
The closure constraint (rather than the topology) forces smaller rings
(e.g., $N=64$ in Fig.~\ref{fig:df}(b))), to form rather dense
globules of blobs. 
%
%(Note that all rings ``show off" with the closure hump
%we had already observed for dilute chains (Fig.~\ref{fig:df}(a))
%when $\dr \approx \Rgg$.)
%

For (asymptotically) long rings there are now two possible scenarios: 
Either all the $\sigma$-lines merge again at large enough distances or they become parallel. 
(We reject crossing of the $\df(\sigma)$-lines as unphysical.)
Unfortunately we are at present unable to decide unambiguously between these two.
We believe it however more likely that all the lines eventually merge and that the 
asymptotic properties of ring polymers are independent of the persistence length $b$
\cite{com:contuniv}. 
Hence, for large distances $\dr > \drstar(\sigma)$ the evolution of $\df(\dr)$ 
becomes again {\em independent} of $\sigma$. It is tempting in view of the scaling
in Fig.~\ref{fig:uv} to set $\drstar = \dtop$ \cite{com:mergedeltar}.

\subsection{Local conformational properties: Contour distance \Rnn}
\label{subsec:G}

We consider finally the average distance 
$\Rnn = \langle (\vec{R}_i-\vec{R}_{i+n})^2\rangle^{1/2}$ 
between the ends of a ring contour segment of length $n$.
Similarly to the discussion above for $S(q)$ we define a differential
fractal dimension $1/\df = d\log(R_n)/d\log(n)$.
Note that $S(q)$ and \Rnsq\ are not just simply Fourier transforms to each other.
They contain different information: 
the $S(q=2\pi/\Rnn)$ monomers of the reference ring within a volume of radius 
$\approx \Rnn$ around an arbitrary monomer of the same ring comprise also monomers 
far away the ring contour, i.e. much larger than $n$. This contribution is 
certainly small for linear chains (hence $S(q) \propto n$ \cite{degennes}), 
but gets more and more important while the rings become more compact 
(with increasing $\sigma$ and $N$).

We have plotted \Rnn\ in the main part of Fig.~\ref{fig:Rn} versus $n \leq N/2$
for various ring lengths mass as indicated in the figure.
Both flexible ($\sigma=0$)
and semi-flexible ($\sigma=3$) are included.
The lines of the stiffer rings are stronger curved.
For $n \ll N/2$ all curves collapse on a chain length independent
(but $\sigma$ dependent) master curve. 
Obviously, in the limit of $n\rightarrow N/2$ the ring closure forces 
\Rnn\ to level off towards \Ree, i.e. $1/\df$ has to vanish.   
The inset shows the differential fractal dimension \df\ for our longest rings 
$N=1024$ for three different persistence lengths. Indeed, \df\ for small $n$ 
increases with $\sigma$. For short flexible rings the fractal dimension takes
more or less the Gaussian value. Consistent with Fig.~\ref{fig:df} we are unable to 
separate clearly the different length scales and observe broad crossover lines.
As mentioned above, \df\ has trivially to diverge for $n \approx N/2$ and
data points $n > N/4$ are influenced by this upper cut-off.
However, we again see that stiffer chains show more compact scaling than 
flexible ones. Qualitatively, the $\sigma=3$-curve is even consistent
with the $\df=4$-window predicted for Gaussian LA  \cite{ORD94}.

In brief, the advantage to use stiff chains is due to an additional 
length scale $\dtop \propto N^0$. Flexible chains, very compact on short length scales 
(see the snapshot on the left of Fig.~\ref{fig:snap}), 
``waste" there a large number \gtop\ of monomers. 
Systems with $N \leq \gtop$ (that is, $\Rgg \ll \dtop$) feature only {\em one} 
characteristic size, their own size, and show CD scaling (with $\alpha=0$). 
Semi-flexible chains, however, using less monomers on short distances below \dtop\
can explore larger distances and become more compact in the scaling sense:
%(i.e., larger $\df$):
They do this by having a lower ceiling on the local density $\rho$ of a single
ring, thereby enhancing the topological interaction with their neighbours
(thus increasing \df).

\section{Dynamics of rings in the melt}\label{sec:dynamics}

The differences in the ring statistics should strongly influence their dynamic 
behaviour. 
But as pointed out in ref.~\cite{MWC96}, there is some experimental evidence 
that the dynamics of rings are similar to their linear counterparts, 
at least up to the largest molecular masses that can readily be obtained. 
Our simulation data, presented below, confirms this.

\begin{figure}
\centerline{\epsfig{file=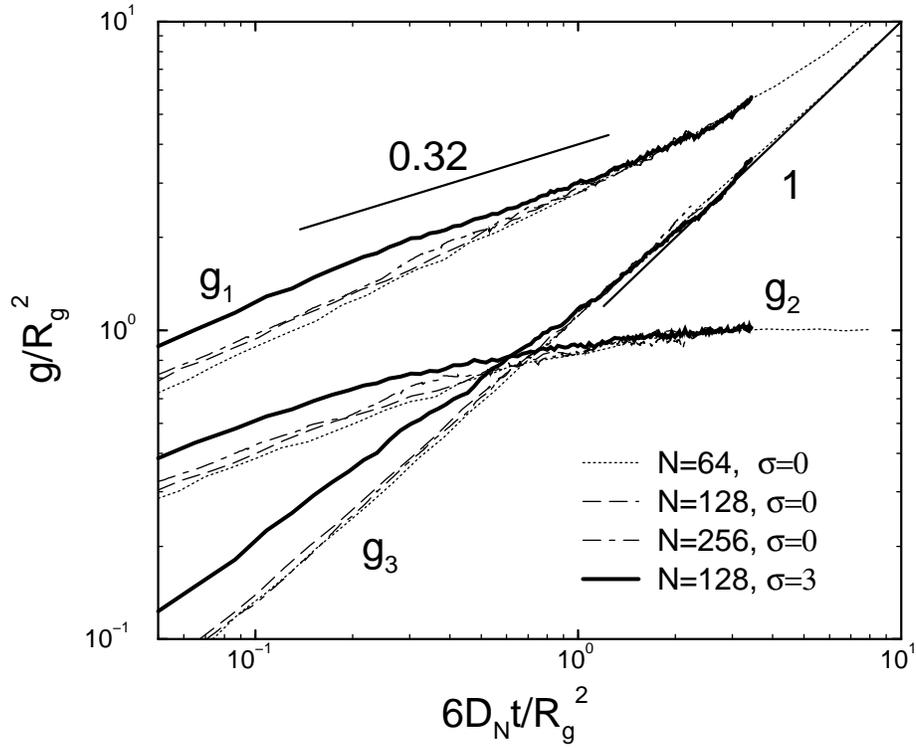,width=120mm,height=100mm}}
\vspace*{1cm}
\caption{Reduced mean-square displacements versus reduced time 
for two rigidities and different chain length as indicated in the figure.
The scaling works reasonably well for flexible chains ($\sigma=0$).
However it is inconsistent with the semi-flexible ($\sigma=3$) configurations.
\label{fig:gsum}}
\end{figure}

\begin{figure}
\centerline{\epsfig{file=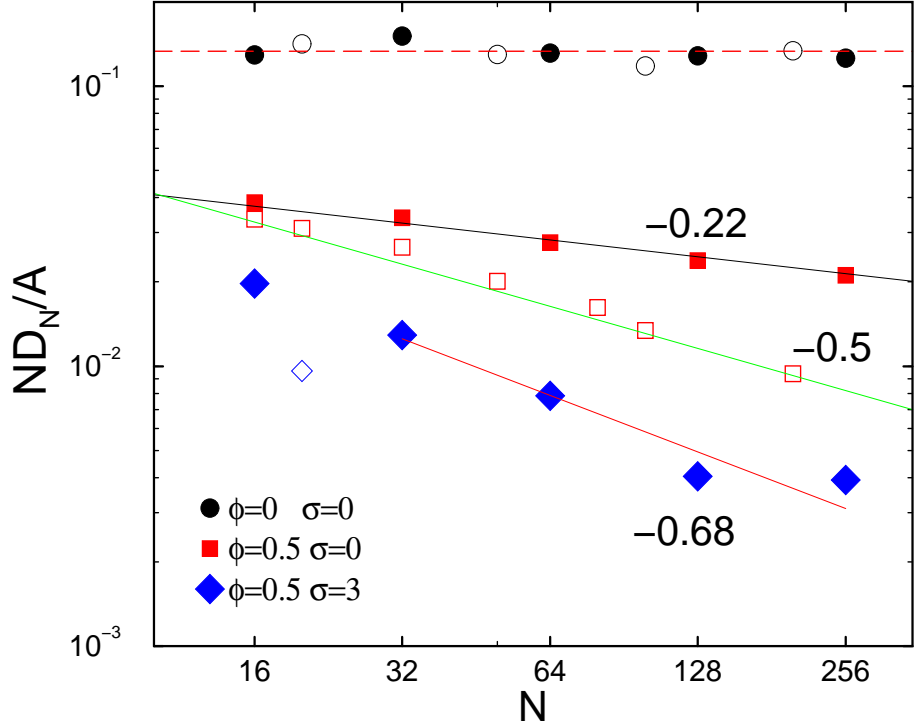,width=120mm,height=100mm}}
\vspace*{1cm}
\caption{Diffusion constants $N \DN$ reduced by the acceptance rate $A$ versus $N$ for 
rings (full symbols) and linear chains (empty symbols).
We consider dilute, flexible ($\sigma=0$) chains (circles)
and flexible ($\sigma=0$, squares) and semi-flexible ($\sigma=3$, diamonds) 
chains in the melt ($\phi=0.5$). The lines correspond to effective power laws.
\label{fig:DN}}
\end{figure}

We characterise the dynamics by measuring three different mean-square displacement functions
describing the motion of monomers 
$g_1(t)=\left< [{\bf R}_n(t) - {\bf R}_n(0) ]^2 \right>$ in the laboratory frame, 
the motion of monomers in the center-of-mass frame of a given ring 
$g_2(t)=\left<[{\bf R}_n(t)-{\bf R}_{c.m.}(t)-{\bf R}_n(0)+{\bf R}_{c.m.}(0)]^2\right>$, 
and the motion of the center of mass itself 
$g_3(t)=\left<[{\bf R}_{c.m.}(t) - {\bf R}_{c.m.}(0)]^2\right>$.
As shown in Fig.\ref{fig:gsum}, the mean square displacements for flexible 
rings collapse onto chain length independent master curves 
when the mean square displacements are rescaled by the radius of gyration \Rgsq, 
and the time by the characteristic relaxation time $\Rgsq/\DN$.
Here \DN\ is the diffusion coefficient (presented in Fig.~\ref{fig:DN}) 
obtained from the asymptotic behaviour of the center of mass motion $g_3(t)/t$
(bold line of slope 1 on the right).  

Closer inspection shows, however, that the collapse is not perfect.
This is in agreement to what was observed by Brown and Szamel \cite{brown}.
In any case, this scaling is not born out for the semi-flexible ($\sigma=3$) systems
of mass $N=128$ depicted in Fig.~\ref{fig:gsum}. This does not come as a surprise
in view of what we have described above in section~\ref{sec:densestatics}
about the scaling of conformational properties.
It is worth noting than in inspection of the differential slope of the 
motion of monomers $g_1(t)$ (not shown) shows (a small, but distinct) 
region with slope 0.32 instead of the classical Rouse-like anomalous
diffusion exponent $\approx 1/2$.

In Fig.\ref{fig:DN} we present the self-diffusion constant of 
linear chains (empty symbols) and rings as a function of 
their size $N$. 
The circles correspond to isolated swollen chains.  
The diffusion constant is the same for linear chains and rings.
It scales like $\DN \sim 1/N$. Our Monte Carlo simulations do not 
of course incorporate hydrodynamic interactions and the dynamics
is expected to be Rouse-like \cite{doi}.
The squares correspond to flexible chains in the melt $\phi=0.5$.
It is important to note that rings are always {\em faster} than their
linear counterpart of same mass. As shown in the figure, 
the diffusion coefficient are well fitted by (effective) power laws: 
$\DN \propto N^{-1.22}$ for rings and
$\DN \propto N^{-1.5}$ for linear chains.
As was emphasised in ref.\cite{MWC96} both linear chains and rings
scale with the size of their correlation hole $\DN\propto R^{-3}$.
The diamonds in Fig.\ref{fig:DN} display the results of semi-flexible rings 
($\sigma=3$) at melt density $\phi=0.5$. The dependence of the diffusion 
constant for the larger ring sizes studied is stronger than for flexible
chains and obeys the apparent relation $\DN \sim N^{-1.68}$.
This is roughly consistent with the data of Brown and Szamel 
who found $D \sim N^{-1.54}$ for large flexible rings \cite{brown}.
Both sets of simulation data are likely to be affected by crossover 
effects.

This observation is in accord with theoretical calculations, which calculate 
the diffusion constant of rings {\em via} the motion of kinks along the contour.
This diffusion of kinks along the molecules is also the dominant relaxation 
mechanism in the reptation of linear chains. In this sense, the dynamics of 
(more and more entangled) rings is the analog of reptation in melts of 
linear molecules.
We do not find any signature of arm retraction in the motion of ring polymers. 
The latter would result in an exponential decrease of the diffusion with ring 
size like in star polymers \cite{stars}.

In the (suspected) LA limit, the ring conformations should possess an 
hierarchical structure. Studying the motion of a single ring in a network of 
(explicitly) fixed obstacles, Obukov {\em et al.} suggested that the outer arms of a LA 
can rearrange much faster than the inner structure \cite{ORD94}.
This results in a broad range of relaxation times. 
We investigated this by monitoring the distribution of mean square displacements
of monomers for flexible rings ($N=256$ and $\sigma=0$) and semi-flexible ones 
($N=256$ and $\sigma=3$) after a time interval 
$\Delta t \approx 2.3 \; 10^5$ MCS. 
This corresponds to a time scale which is shorter than the diffusion time 
$\DN/\Rgg^2$ of the ring. Note that both curves correspond to roughly the same 
time period {\em and} the same mean value of the monomer displacement.
The center of mass diffusion constant differs, however, by an order of 
magnitude. Hence, the major difference in the dynamic behaviour does not stem 
from a difference in the local dynamics (e.g.\ a slight decrease of the 
acceptance ratio of the Monte Carlo moves due to the additional bond angle 
potential), but reflects the interplay between conformations/topology and 
the dynamics. Notably, the monomer mean square displacements for flexible
rings are Gaussian distributed, while the distribution of semi-flexible 
systems is non-Gaussian. This is consistent with simulations of rings in 
a network of fixed obstacles \cite{ORD94}.

From these different probes of the ring dynamics we conclude that the 
dynamics for flexible and semi-flexible chains are qualitatively different,
at least for the chain lengths we were able to probe.
There appears to be clear evidence for the occurrence of a second length
and time scale for our semi-flexible chains (which have $R \ge \dtop$), 
but not for our flexible ones (which have $R \le \dtop$).
Qualitatively we conclude again that stiffer rings are closer to the LA limit
than their flexible counterparts. However, even for the largest ring size and 
stiffness our data are still affected by the small number of arms and we expect 
pronounced finite size effects to the asymptotic behaviour for large rings 
in the melt. An investigation of the dynamics of even larger rings ($N>256$) is unfortunately 
beyond our computational facilities at present.

\section{Conclusions}\label{sec:summary}

In summary, we have presented extensive Monte Carlo simulations of semi-dilute 
solutions and melts of flexible and semi-flexible ring polymers. The rings are 
neither knotted with themselves nor concatenated with each other. 

In order to have a reference with systems where the non-linkage constraint
does not play a r\^ole we have briefly considered dilute solutions of rings.
They appear to be extremely similar to their linear counterparts, both with 
respect to statical and dynamical properties. The (remaining) topological 
unknottedness constraint seems not to be pertinent \cite{deutsch}, at least not for the 
chains sizes ($N=1024$) we have considered here. The ring extension scales 
like $R \sim N^{\nu}$ with $\nu \approx 0.59$. This was also made evident 
from the study of the differential fractal dimension $\df(q)$ obtained from 
the static structure factor $S(q)$. 
Dilute rings do not differ from linear chains with regard to the influence of
finite persistence length (at least for large enough rings).
The dynamics is Rouse-like $\DN \sim N^{-1}$, 
because our Monte Carlo simulations ignore hydrodynamical interactions 
(``free draining limit''). 
In short, from the practical point of view topology is irrelevant.

This is dramatically different in the high density limit ($\phi=0.5$) with 
strongly overlapping entangled rings where topological constraints tend to 
squeeze the rings into relatively compact objects.
We have varied the stiffness of the rings so as to tune the overlap between 
different rings. Due to the rather compact structure of the molecules in the 
melt, increasing the stiffness is much more efficient than increasing the ring size. 
Essentially, stiffer chains ``waste" less monomers on short distances and have more 
monomers left to meander through the topological constraints imposed by neighbouring rings.
Indeed the {\em effective} Flory exponent $\nu(N)$ obtained from the high chain length 
behaviour of chain diameter and radius of gyration shows a strong effect with regard
to the persistence length, decreasing from $\nu \approx 0.4$ for flexible chains 
($\sigma=0$) to $\nu\approx 1/3$ for our stiffest systems ($\sigma=3$). 
Chain stiffness allows a reduction in $\nu$ to more ``compact'' values by {\em increasing}
(at a given $\nu$) the overlap parameter $\nrover \propto b^3 N^{3\nu-1}$.

%Scaling
%
Rings with topological constraints do not follow the classical one parameter scaling 
with $\phi/\phistar$ for linear chains where the size of the (dilute) chain of mass $N$ 
and stiffness $\sigma$ sets the only relevant length scale \cite{degennes}. 
This is in disagreement to the fundamental assumption of the CD picture\cite{CD86}. 
In order to scale the chain length $R=R(N,\sigma)$ we were
forced to assume an additional chain length independent length scale 
$\dtop \propto \gtop^{\nu_1}$. 
Supposing a weak $\sigma$-dependence of \dtop\ and choosing $\nu_1$ self-consistently
this yields a satisfactory data collapse and $\nu_1 \approx 1/2$ as demonstrated in 
Fig.~\ref{fig:uv}. 
For small chains, $R \ll \dtop$, the topological interactions are weak.
The rings resemble their linear counterparts and behave effectively like closed 
Gaussian chains of blobs. This regime is consistent with the $\alpha \rightarrow 0$
limit of the CD scenario, i.e. due to the non-linkage constraint the free energy
of a reference ring increases only like $\log(\nrover)$.
Rings larger than \dtop\ appear to be governed by the topological interactions
and are more compact (in the scaling sense of smaller $\nu$). 
They are well characterised by an effective Flory exponent $\nu_2 = 1/3$.
This scaling scenario is broadly consistent with the concept of  
{\em lattice animals} (LA) within a network of topological obstacles
created in a self-consistent manner by surrounding rings\cite{ORD94,KN96}. 
These LA appear to be made of locally Gaussian chain pairs at short distances.

Our (more ambitious) discussion of the differential fractal dimension $\df$ 
(obtained from the static structure factor $S(q)$ and/or 
the spatial distance $\Rnn(n)$ between a contour segment of length $n$)
reveals very rich and broad crossover effects:
excluded volume $\xi$ versus persistence length \xikuhn,
excluded volume versus ring closure, 
ring size $R$ versus topological length scale \dtop.
As shown in Fig.~\ref{fig:df} and Fig.~\ref{fig:Rn}, it is not possible to separate 
the different length scales unambiguously and to disentangle their physics
\cite{com:logcorr}.
In view of the (restricted) range of parameters ($N\leq 1024,\sigma \leq 3, \phi=0.5$) 
we are able to simulate, this does not come as a surprise.
Much more surprising is the success and the simplicity of the scaling 
scenario of  Fig.~\ref{fig:uv} for the global chain $R$ size described 
in the paragraph above.
There, all the intricate short range physics was cast in {\em one} effective exponent 
$\nu_1$ for all chain and persistence lengths. 
It just turns out that the Gaussian value $\nu_1=1/2$
(i.e. $\alpha \rightarrow 0$) fits particular well the data. 
Similarly, while $\nu_2=1/3$ is certainly the asymptotic value, this does
not exclude the possiblity of an intermediate window with $\nu_2=1/4$.
Indeed the differential fractal dimensions for our largest and stiffest 
configurations clearly exceed $\df=3$. This is in favour of an ultra-compact
transient which should then eventually become evident also in the ring sizes
for even larger chain lengths $N$ than we are at present able to simulate.

Additional evidence for the crossover to a strongly entangled regime 
characterised by an additional length scale \dtop\ comes from our brief
investigation of the ring dynamics.
The scaled time dependence of the monomer displacements differs from the 
master curve in the non-entangled regime, and the monomer displacements at 
times smaller than the relaxation time are non-Gaussian distributed as expected 
for LA \cite{ORD94}. 
The diffusion constant for chains in the melt scales like $\DN \sim N^{-1.22}$
for flexible chains. This decrease to $\DN \sim N^{-1.68}$ for semi-flexible
systems ($\sigma=3$). Again these exponents are presumably only effective values 
due to a broad crossover between unentangled and entangled regime. 
The similarity of the dynamics of melts of rings and linear chains suggests that our observations 
for rings might also be pertinent to the dynamics of linear chains. Indeed, it is tempting to relate
the topological constraints which lead to the lattice animal behavior for rings to the entanglements in 
linear chains. While the topological interactions do not influence the static conformations of linear 
chains, however, rings offer the additional possibility to investigate the effect of topology in the 
static behavior.

% Final Paragraph ...
In any case, our simulation data cover only the onset of LA behaviour 
and our estimates for the scaling functions of the conformational statistics 
and the dynamics are likely to be subjected to corrections due to the very 
small number of arms. 
In future we plan to corroborate further the discussion of the dynamical properties
(increasing the number of statistical segments) and to investigate the static and 
dynamical scaling properties with regard to the monomer density $\phi$ \cite{MWC99b}.

\subsection*{Acknowledgement}

M.M. has benefitted from interesting discussion with W.~Paul, K.~Binder, 
and K.S.~Schweizer. He thanks EPCC for very kind hospitality and financial 
support during his stay as a TRACS visitor and J.-C.~Desplat for 
visualisations of the ring conformations (Fig.~\ref{fig:snap}). 
A generous grant of CPU time on the CRAY T3E computers at the EPCC in Edinburgh and 
the HLR Stuttgart as well as partial financial support by the DFG under grant Bi314-17 
are gratefully acknowledged.
J.P.W. acknowledges stimulating discussions with S.P.~Obukov which triggered 
this study. Moreover he is indebted to him for Russian driving lessons in 
California. Finally he would like to thank C.~Gay and J.-L.~Barrat for detailed
comments concerning the topological constraints.

%%%%%%%%%%%%%%%%%%%%%%%%%%%%%%%%%%%%%%%%%%%%%%%%%%%%%%%%%%%%%%%%%%%%%%%
\newpage

\begin{table}[htbp]
\begin{tabular}{|c|c|c|c|c|c|c|} \hline
$\sigma$&\costh& $l$   & \beL    & \beR   &  $A$  & $N \DN$ \\ \hline
0       &-0.193& 2.736 & 3       & 1.55   &  0.254& 0.032  \\
1       &-0.394& 2.724 & 3.32    & 1.73   &  0.224& 0.026  \\
2       &-0.554& 2.706 & 3.94    & 1.89   &  0.193& 0.021  \\
3       &-0.668& 2.703 & 4.24    & 2.24   &  0.165& 0.019  \\ \hline
\end{tabular}
\vspace*{1cm}
\caption{
Persistence length dependent properties for dilute ($\phi=0$) rings and linear chains:
Mean cosine of the bond-bond angle \costh, 
mean bond length $l$, effective bond length of linear chain \beL\ 
and rings \beR, acceptance rate $A$ and the diffusion constant $N \DN $. 
These values characterise the asymptotic behaviour of long linear chains as well as rings.
\label{tab:dilutesigma}}
\end{table}
 
\begin{table}[htbp]
\begin{tabular}{|c|c|c|c|c|c|} \hline
$\sigma$ &\costh & $l$   & \beL  & $A$    & $\nu_{eff}$ \\\hline
0        &-0.106 & 2.632 & 3.2   & 0.1529 & 0.41 \\
1        &-0.348 & 2.614 & 3.7   & 0.1329 & 0.39 \\
2        &-0.544 & 2.602 & 4.5   & 0.1136 & 0.36 \\
3        &-0.676 & 2.593 & 5.3   & 0.0983 & 0.33 \\ \hline
\end{tabular}
\vspace*{1cm}
\caption{
Persistence length dependent properties for rings and linear chains
at high volume fraction $\phi=0.5$:
Mean cosine of the bond-bond angle \costh, mean bond length $l$, 
effective bond length \beL\ for linear chains, acceptance rate $A$
and the differential Flory exponent $\nu_{eff}$ fitted over the four largest 
chains available (see Fig.~\ref{fig:RsN}).
\label{tab:densesigma}}
\end{table}

\begin{table}[htbp]
\begin{tabular}{|c|c|c|c|c|c|c|c|c|c|c|} \hline
           &  \multicolumn{3}{c|}{$\sigma=0$} & 
              \multicolumn{2}{c|}{$\sigma=1$} & 
              \multicolumn{2}{c|}{$\sigma=2$} & 
              \multicolumn{3}{c|}{$\sigma=3$}\\ \hline
$N$ & \Rgsq & \Resq & $N\DN 10^3$ & \Rgsq & \Resq & \Rgsq & \Resq & \Rgsq & \Resq & $N\DN 10^3$ \\ \hline
% 0.3  &16  &  14.9  & 49.5  & 18.5  & 62.2  & 23.4  &  82.0  & 28.2  & 103  \\
%     &32  &  30.5  & 98.1  & 38.1  & 121   & 50.3  &  161   & 65.8  & 219  \\
%     &64  &  60.0  & 187   & 74.4  & 227   & 98.1  &  297   & 130   & 398  \\
%     &128 &  117   & 355   & 143   & 426   & 187   &  552   & 244   & 716  \\ \hline
%
16  &  12.9  & 42.3  & 5.87   &16.4  & 54.1  & 21.3  &  73.9  & 26.2  & 95.3 & 1.94 \\
32  &  25.7  & 80.6  & 5.18   &32.9  & 102   & 44.7  &  141   & 60.2  & 198  & 1.23\\
64  &  49.3  & 150   & 4.21   &63    & 189   & 84.5  &  251   & 117   & 353  & 0.77\\
128 &  92.2  & 275   & 3.65   &115   & 337   & 150   &  430   & 203   & 579  & 0.4\\
256 &  169   & 492   & 3.23   &204   & 581   & 250   &  700   & 330   & 910  & 0.38\\
512 &  297   & 856   &        &343   & 978   & 420   &  1157  & 490   & 1315 & \\ 
1024&  514   & 1474  &        &539   & 1680  & 620   &  1784  & 710   & 2140 &  \\ \hline
\end{tabular}
\vspace*{1cm}
\caption{
Ring radius of gyration \Rgg, diameter \Ree\ and (reduced) diffusion constant $N\DN 10^3$
for different ring sizes $N$ and stiffnesses $\sigma$ at $\phi=0.5$.\label{tab:dense}}
\end{table}

%%%%%%%%%%%%%%%%%%%%%%%%%%%%%%%%%%%%%%%%%%%%%%%%%%%%%%%%%%%%%%%%%%%%%%%%%%%%%
\newpage

\newpage
\begin{figure}
%\centerline{\epsfig{file=snap.eps,width=120mm,height=100mm}}
\vspace*{1cm}
\caption{Configurational snapshots of rings of mass $N=256$:
dilute flexible ring in the middle, 
flexible ($\sigma=0$) ring in the melt ($\phi=0.5$) on the left
and corresponding semi-flexible ($\sigma=3$) coil on the right.
The overlap has increased by factor three between the flexible and
the semi-flexible chain in the melt. 
To obtain a similar effect by increasing the chain mass
we would have to increase $N$ by at least two orders of magnitude.
\label{fig:snap}}
\end{figure}

\begin{figure}
\centerline{\epsfig{file=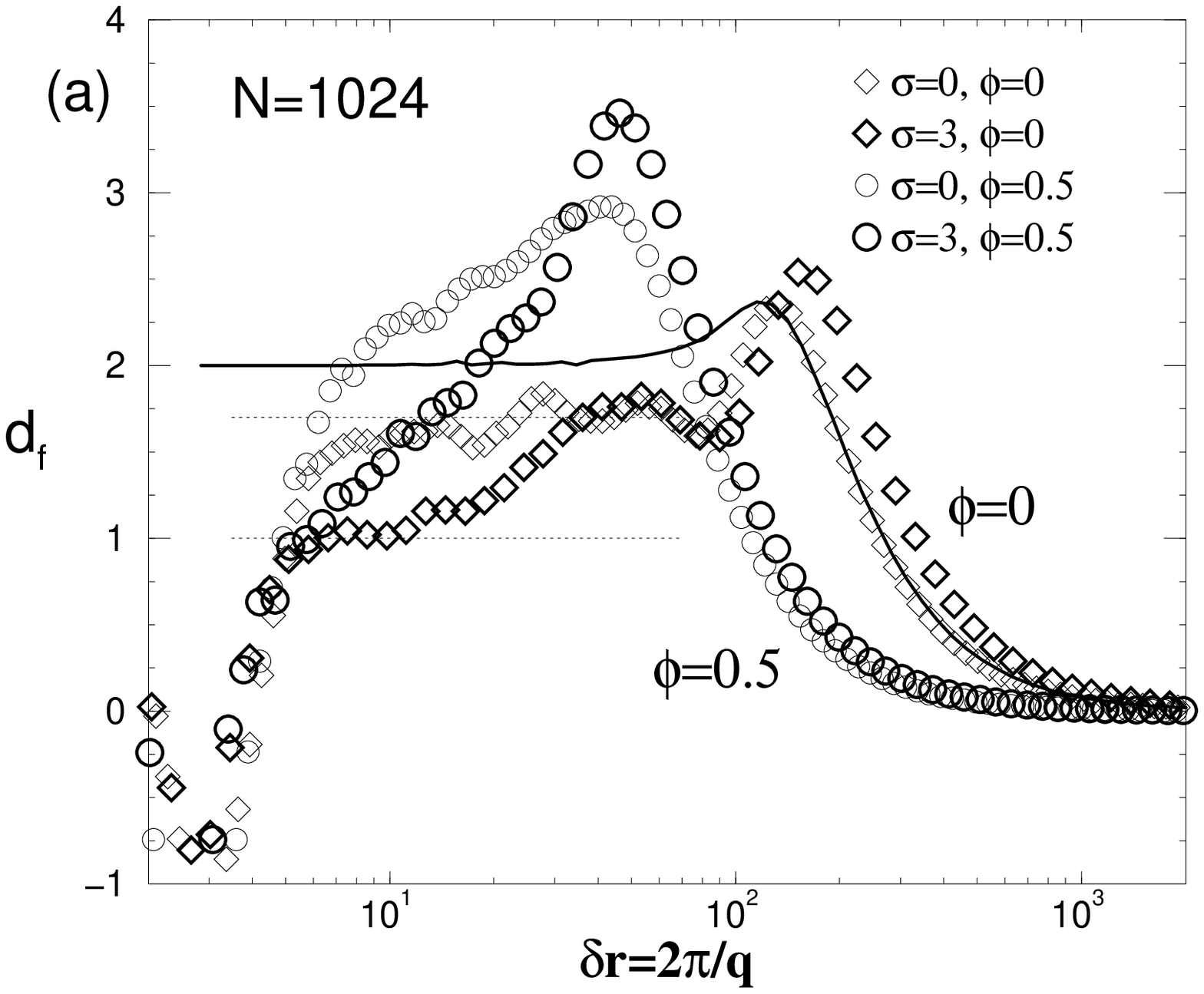,width=120mm,height=90mm}}
\centerline{\epsfig{file=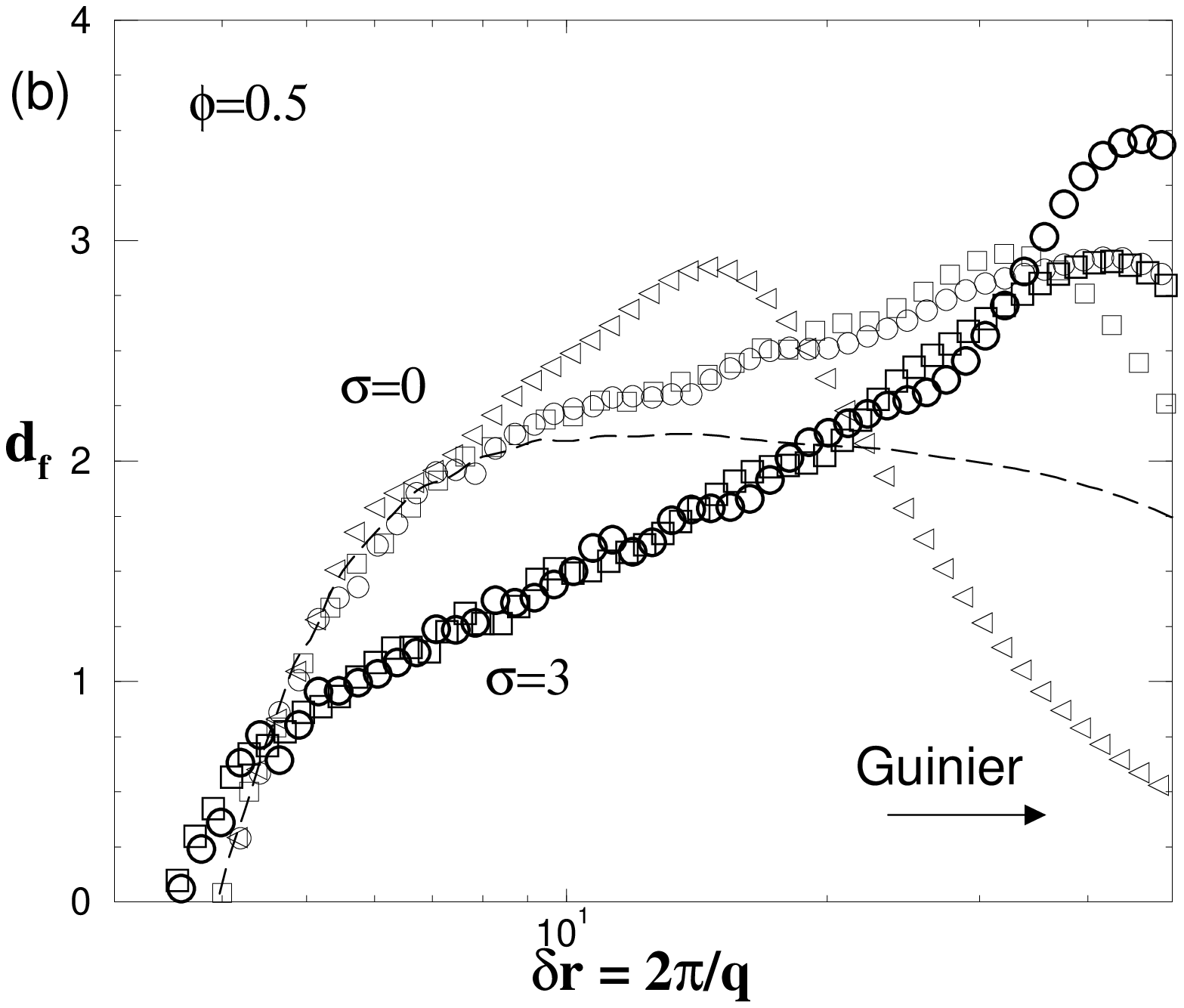,width=120mm,height=90mm}}
\vspace*{0.1cm}
\caption{Differential fractal dimension $\df(q)= - d\log(S(q))/d\log(q)$ 
versus $\dr=2\pi/q$.
(a) Dilute and molten rings of mass $N=1024$.
The line indicates an ideal Gaussian ring of same radius of gyration
as a flexible and dilute ring.
(b) Rings in their melt 
($\phi=0.5$) for $N=1024$ (circles), $N=512$ (squares), $N=256$ (diamonds)
and $N=64$ (triangle). 
%Flexible chains are very compact on short scales.
%The semi-flexible chains waste less monomers on short distances.
%Presumably all lines merge again at large distances.
The dotted line is a comparison with linear chains ($N=256,\phi=0.5$).
Within the first blob ($\dr < \xi(\sigma)$) linear chains and rings 
behave identically.
\label{fig:df}}
\end{figure}

\begin{figure}
\centerline{\epsfig{file=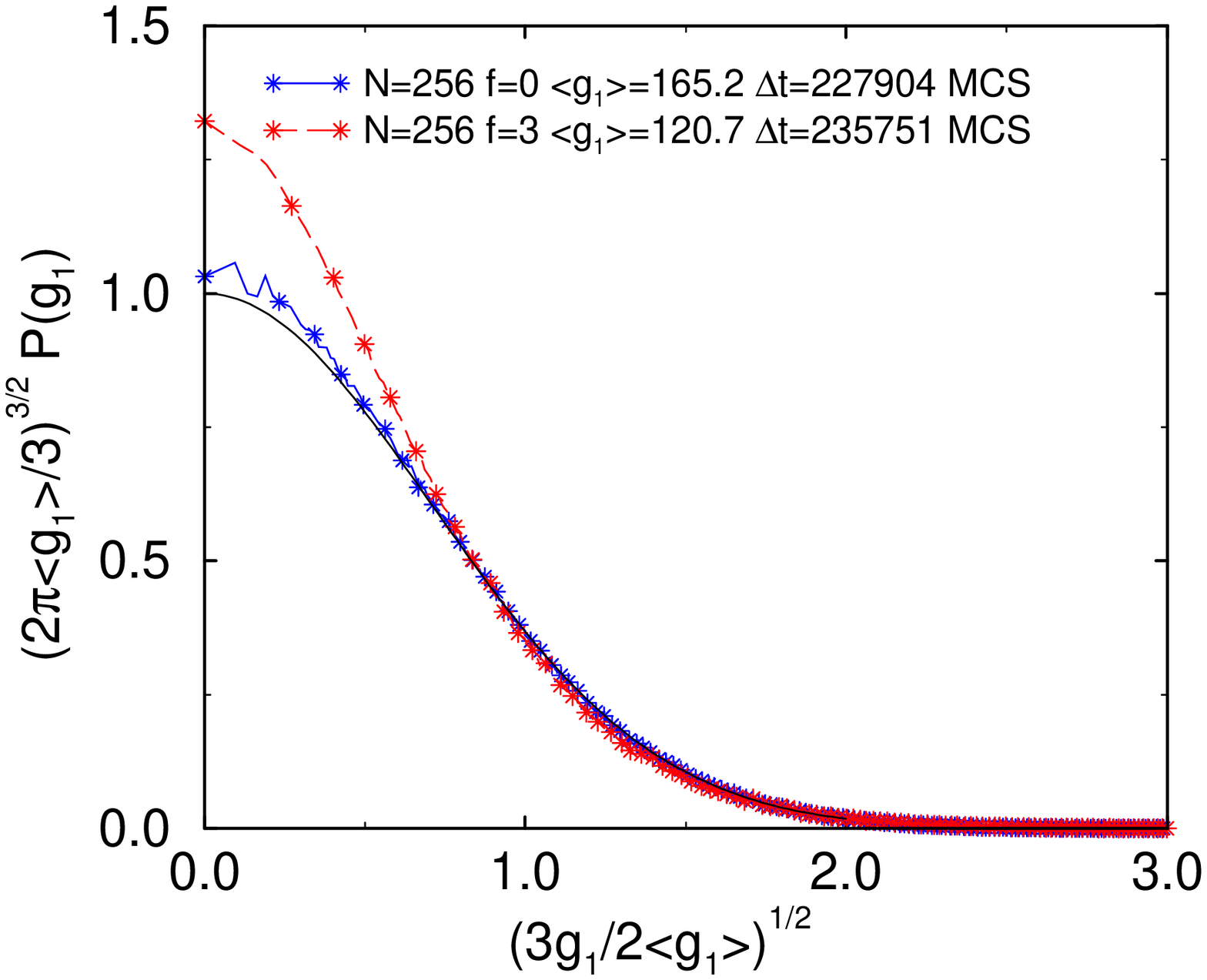,width=120mm,height=100mm}}
\vspace*{1cm}
\caption{Probability distribution of mean-square displacements of monomers
at short times. 
Flexible rings ($\phi=0.5,\sigma=0$):
Gaussian distribution of monomer displacements at $6D\Delta t/R_g^2=0.42$.
Semi-flexible rings ($\phi=0.5,\sigma=3$):
deviations from the Gaussian behavior at $6D\Delta t/R_g^2=0.02$ indicate 
abundance of slow monomers (presumably) within the ``trunk'' of the 
lattice animal tree.
The solid line represents the Gaussian distribution.
\label{fig:probg1}}
\end{figure}


\begin{thebibliography}{99}
\bibitem{com:experiments}
Unknotted and unconcatenated ring polymers have actually been made and
investigated experimentally.
The reader will find in ref.~\cite{MWC96} a detailed account of results and
technical difficulties related to ring synthesis and contamination with linear rings. 
We do not wish to repeat this here. (See also ref.~\cite{lrp}.) 
All these studies focus on dynamical properties. 
No experimental study of the radius of gyration of rings in melt appears to have 
been made till now  --- despite the fact that any reasonable description of dynamics 
requires a good understanding of conformational properties.
\bibitem{com:estimateAlgo}
These estimates depend strongly on the further evolution of the effective 
Flory exponent $\nu(N)$ discussed below. However,
the balance remains strongly favourable even if the computational overhead 
(to calculate the additional potential), the reduced acceptance rate $A$ 
(see Tab.~\ref{tab:densesigma}) and (more importantly) 
the slower diffusion (see Fig.~\ref{fig:DN}) are taken into account.
\bibitem{com:valuegauss} 
We have also tried values based on the effective bond lengths for Gaussian chains from 
ref.~\cite{WPB92} quoted in Tab.~\ref{tab:densesigma}. 
The results are however virtually identical.
\bibitem{com:needle}
Let us assume that the ring consists of uncorrelated needle-like arms of 
a fixed length $\Delta$. Two segments on the same arm are either parallel
or anti-parallel; the directions of different arms are completely uncorrelated.
For this crude model the bond-bond correlation function 
$\langle \vec{e}_i\vec{e}_{i+n} \rangle$
decays linearly from 1 at $n=0$ to $-1/2$ at $n=\Delta$, and then increases linearly to
$0$ at $n=2\Delta$. For larger distances $n$ along the chains there are no correlations by
construction. This motivates that a minimum in the bond-bond correlation function is associated
with a back-folding and the position of the minimum is an estimate for the arm length. Certainly,
there is a distribution of arm lengths and the correlation along an arm is not perfect. This 
will broaden the minimum. More important, the above argument neglects the hierarchical structure
of lattice animals and only captures local correlations.
\bibitem{com:warninglinear}
Note that even for linear chains of length $N=256$ we do not 
get a perfect Gaussian plateau, but rather a flattish hump
(see Fig.~\ref{fig:df}(b)). This provides an important warning.
Even for the linear chains it is tricky to reach the asymptotic limit
for the differential fractal dimension.
\bibitem{com:mergedeltar}
Accepting all lines to merge eventually, we have to ask how.
The simplest suggestion is that they all merge together at one point $\drstar$. 
Then no persistence length, e.g. the $\df(\sigma=0)$-line, is singled out.
Otherwise one may characterise each persistence length by the point $\drstar(\sigma)$ 
where it joins the flexible ring line $\df(\sigma=0)$.
\bibitem{com:logcorr}
Not to mention logarithmic corrections due to weak topological interactions
in the $\alpha\rightarrow 0$ limit of the CD regime.
\end{thebibliography}
\end{document}